\documentclass[useAMS,usenatbib,usedcolumn]{mnras}

\pdfoutput=1
\usepackage{graphicx}
\usepackage{xcolor}
\usepackage{booktabs}
\usepackage{wrapfig}

\def\lesssim{\la}
\def\gtrsim{\ga}

\usepackage{flushend}

\title[Velocity function and alternative dark matter]{Hints against the cold and collisionless nature of dark matter from the galaxy velocity function}

\author[Schneider et al.]{Aurel Schneider$^{1}$, Sebastian Trujillo-Gomez$^{2}$, Emmanouil Papastergis$^{3}$,
\newauthor{Darren S. Reed$^{2,4}$, and George Lake$^{2}$}\\
\\
{$^1$Institute for Astronomy, Department of Physics, ETH Zurich,
Wolfgang-Pauli-Strasse 27, 8093, Zurich, Switzerland}\\
{$^2$Institute for Computational Science, University of Zurich, Winterthurerstrasse 190, 8057 Zurich, Switzerland}\\
{$^3$Kapteyn Astronomical Institute, University of Groningen, Landleven 12, Groningen NL-9747AD, The Netherlands}\\
{$^4$$S^{3}IT$, University of Zurich, Winterthurerstrasse 190, 8057 Zurich, Switzerland}\\
{Email: aurel.schneider@phys.ethz.ch}}

\begin{document}

\label{firstpage}
\maketitle

\begin{abstract}
The observed number of dwarf galaxies as a function of rotation velocity is significantly smaller than predicted by the standard model of cosmology. This discrepancy cannot be simply solved by assuming strong baryonic feedback processes, since they would violate the observed relation between maximum circular velocity ($v_{\rm max}$) and baryon mass of galaxies. A speculative but tantalising possibility is that the mismatch between observation and theory points towards the existence of non-cold or non-collisionless dark matter (DM). In this paper, we investigate the effects of warm, mixed (i.e warm plus cold), and self-interacting DM scenarios on the abundance of dwarf galaxies and the relation between observed HI line-width and maximum circular velocity. Both effects have the potential to alleviate the apparent mismatch between the observed and theoretical abundance of galaxies as a function of $v_{\rm max}$. For the case of warm and mixed DM, we show that the discrepancy disappears, even for luke-warm models that evade stringent bounds from the Lyman-$\alpha$ forest. Self-interacting DM scenarios can also provide a solution as long as they lead to extended ($\gtrsim 1.5$ kpc) dark matter cores in the density profiles of dwarf galaxies. Only models with velocity-dependent cross sections can yield such cores without violating other observational constraints at larger scales.
\end{abstract}



\section{Introduction}
Observations of the abundance and structure of dwarf galaxies have the potential to probe the particle nature of dark matter (DM). This is because effects from DM free-streaming or from (self-)interactions have an impact on structure formation at the smallest observable scales.

There are potential inconsistencies between small-scale observations and the standard model of $\Lambda$CDM \citep[based on the observations of the {\tt Planck} satellite,][]{Planck:2015xua}. First of all, haloes predicted by gravity-only simulations greatly outnumber observed galaxies. This long-standing discrepancy has been established for both the Milky-Way satellites \citep{Klypin:1999uc,Moore:1999nt} as well as nearby isolated galaxies \citep[e.g.][]{Tikhonov:2009jq,Zavala:2009ms} and is usually referred to as the {\it over-abundance} (or {\it missing satellite}) problem. Second, observations of rotation velocities from stars and gas point towards very shallow inner density profiles of small haloes in strong contrast to predictions from gravity-only simulations. This is generally known as the \emph{cusp-core} \citep[e.g.][]{deBlok:2009sp} or the \emph{too-big-to-fail} problem \citep[TBTF,][]{BoylanKolchin:2011de,Papastergis:2014aba}, depending on the context.

The main difficulty with these problems of small-scale structure formation is the fact that they are based on predictions from gravity-only simulations, ignoring any potential effects from baryonic physics. Indeed, it is expected that photo-evaporation from UV sources during reionisation can expel gas from small haloes, effectively preventing star formation and reducing the number of observable dwarf galaxies \citep[e.g.][]{Gnedin:2000uj,Okamoto:2008sn}. More recently, it was realised that supernova feedback is energetic enough to reshape the inner parts of halo profiles making them considerably shallower \citep{Governato:2012fa}. However, the details of how feedback affects the halo profile are still under debate. For example, \citet{Onorbe:2015ija} and \citet{Read:2015sta} show that the core size depends on the details of the star formation history, while \citet{DiCintio:2013qxa} and \citet{Fitts:2016usl} connect the core size to the dwarf's stellar-to-halo mass ratio. Other papers \citep[e.g.][]{Sawala:2015cdf,Fattahi:2016nld} point out that the presence of cores depends on the feedback implementation and might not be required to recover the observations.

Despite the ongoing debate about the efficiency of baryon-induced feedback mechanisms, they are generally assumed to be the most likely explanation for the dwarf abundance and structure problems. However, because of our poor understanding of sub-grid effects in hydrodynamical simulations, it has so far been impossible to verify these assumptions from first principles.

A very useful statistic that simultaneously probes both the abundance and the structure of galaxies is the velocity function (VF), i.e. the number density of galaxies as a function of their observed rotation velocity. The VF offers a direct link between observations and theory because the rotation velocities of galaxies act as a tracer of the halo gravitational potential. Since the galaxy velocity function is sensitive to both the abundance and the inner structure of haloes, any model that predicts the observed VF is also likely to solve the overabundance, the cusp-core, and the TBTF problems. This makes the velocity function an ideal probe of structure formation.

In a recent study, \citet{Klypin:2014ira} compiled the velocity function of the local 10 Mpc around the Milky Way using rotation velocities ($v_{\rm rot}$) predominantly based on spatially unresolved HI line widths. When compared to the VF from gravity-only simulations, they found a discrepancy in the abundance of galaxies at $v_{\rm rot}<80$ km/s which increases towards lower velocities. However, this comparison is based on the assumption that $v_{\rm rot}$ is a good approximation of the maximum circular velocity ($v_{\rm max}$) of the halo, which is not guaranteed.

In a recent paper, we investigated the relation between the measured rotation velocity of a galaxy and the maximum circular velocity of the halo that hosts the galaxy, and we studied how the resulting $v_{\rm max}$ based velocity function is affected by baryonic processes \citep[][henceforth TG16]{Trujillo-Gomez:2016pix}. Our main findings were the following: (i) the bias between $v_{\rm rot}$ and $v_{\rm max}$ is not large enough to significantly reduce the mismatch between the observed VF and the one predicted by gravity-only simulations of $\Lambda$CDM; (ii) while baryonic processes are able to reduce the theoretical abundance of galaxies and alleviate the over-abundance problem, they cannot completely solve it without simultaneously violating the observed relation between baryon mass and $v_{\rm max}$ (i.e. the $v_{\rm max}$ baryonic Tully-Fisher relation).


This paper builds upon TG16 and investigates the tantalising possibility that the mismatch between the observed and predicted galaxy velocity function is caused by the underlying particle properties of dark matter. Both non-cold and non-collisionless DM models could provide more natural solutions, as they suppress the amplitude of matter perturbations and/or alter the halo density profiles. As representative examples, we focus on the effective scenarios of warm, mixed (i.e. warm plus cold) and self-interacting DM.

The paper is structured as follows: In Sections~\ref{sec:problem} and \ref{sec:theory} we give a brief summary of the results obtained in TG16, and we present our theoretical model of the DM halo velocity function based on the the extended Press-Schechter approach. Sections~\ref{sec:WDM}, \ref{sec:MDM}, and \ref{sec:SIDM} present the results for warm, mixed, and self-interacting DM. We examine how these DM scenarios affect the halo profiles, the maximum circular velocities, and finally the VF. In Section~\ref{sec:othercandidates} we discuss qualitatively other potential DM particle scenarios. Our results are summarised in Section~\ref{sec:conclusions}.


\section{Setting up the problem}\label{sec:problem}
The number density of galaxies as a function of rotational velocities -- i.e. the velocity function (VF) -- is a very useful observational quantity relating information about galaxy abundance with the underlying halo potentials. This allows to compare theory with observations without detailed knowledge about the galaxy formation efficiency. In this section, we recap the results from TG16, summarising the procedure for obtaining maximum circular velocities ($v_{\rm max}$) from observed HI line widths ($w_{50}$) and how this affects the shape of the VF.

\begin{figure*}
\center{
\includegraphics[width=.328\textwidth,trim={0.6cm 0.7cm 5.07cm 0.4cm}]{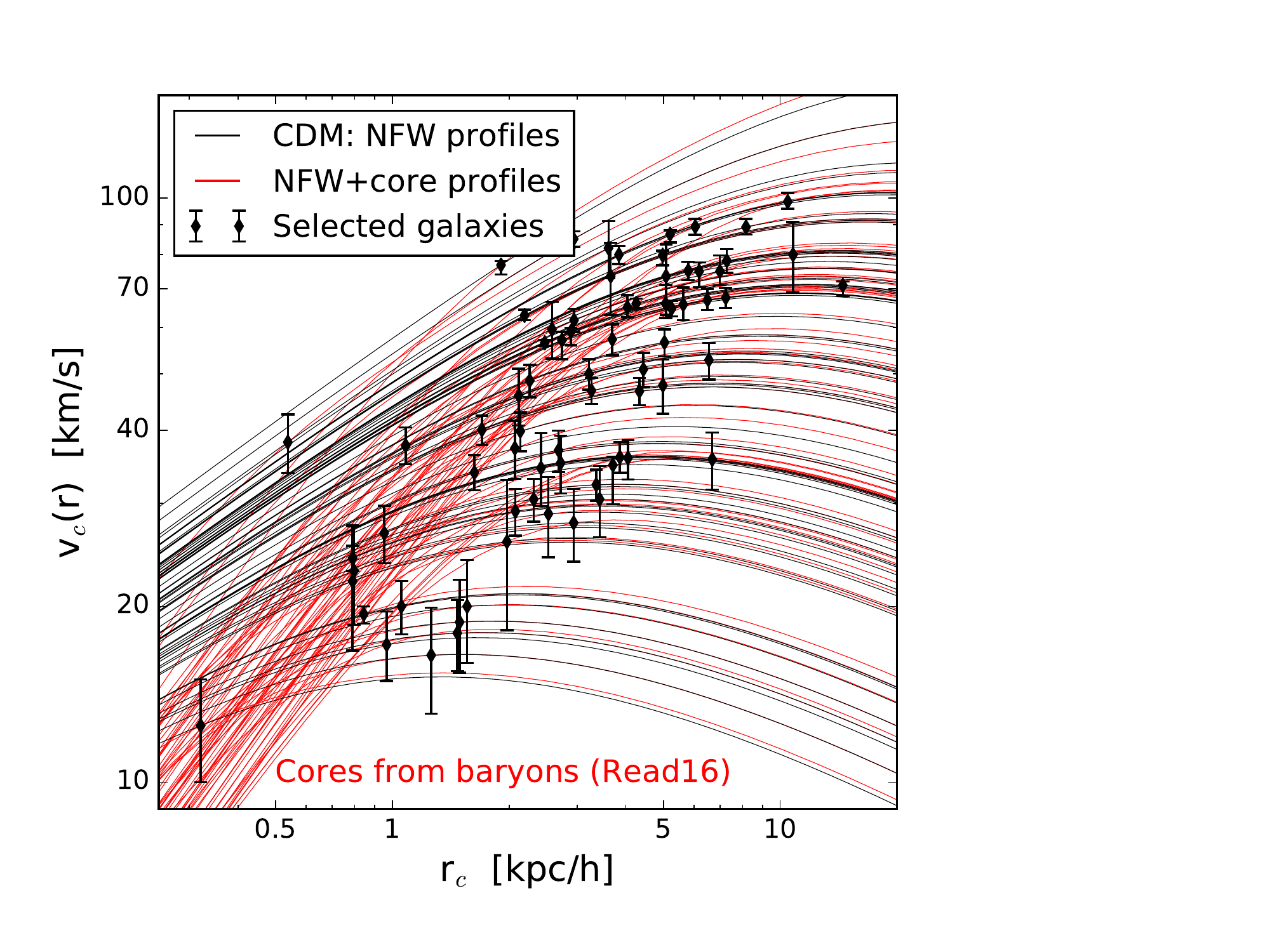}
\includegraphics[width=.328\textwidth,trim={0.6cm 0.7cm 5.07cm 0.4cm}]{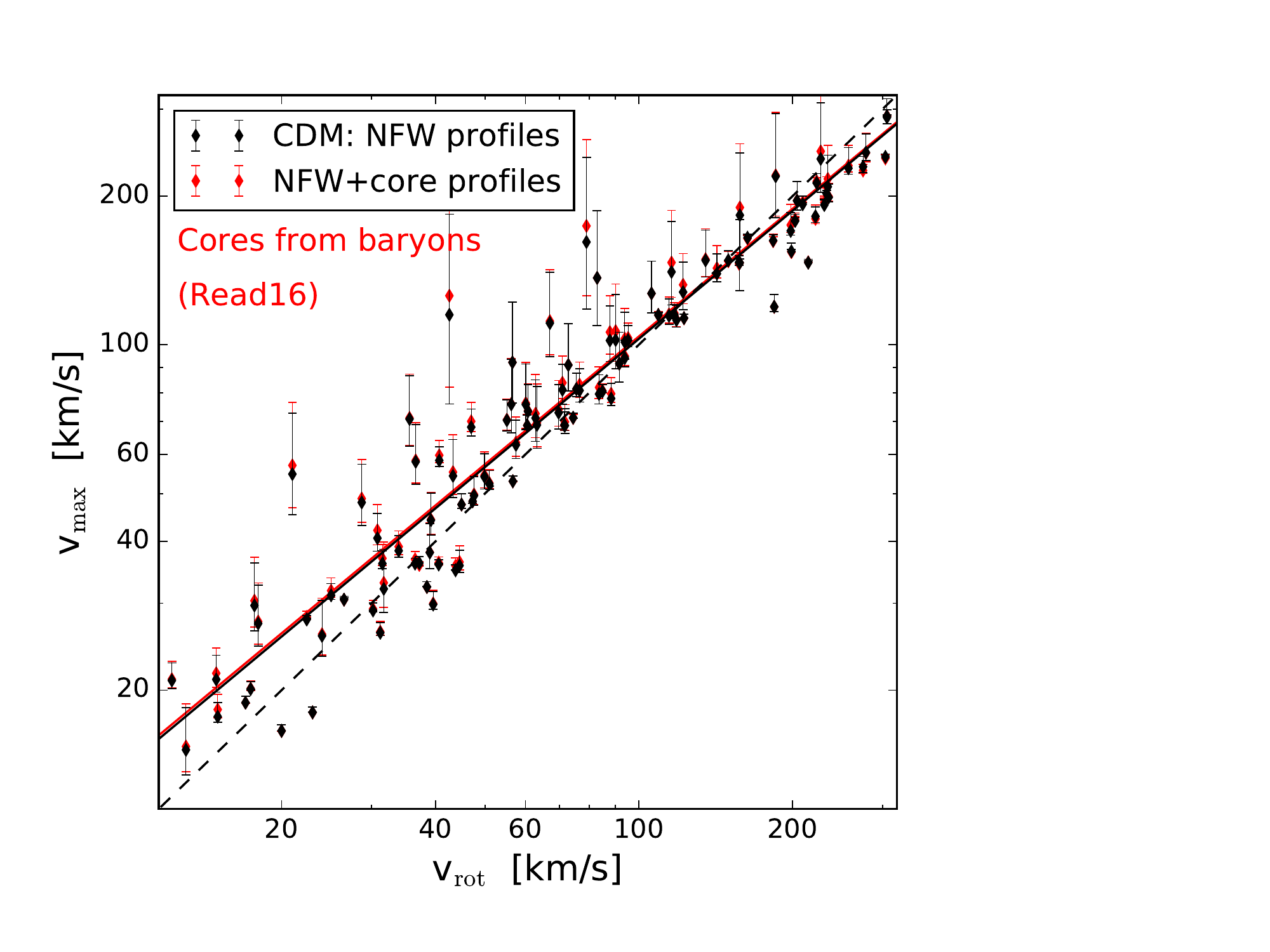}
\includegraphics[width=.328\textwidth,trim={0.6cm 0.7cm 5.07cm 0.4cm}]{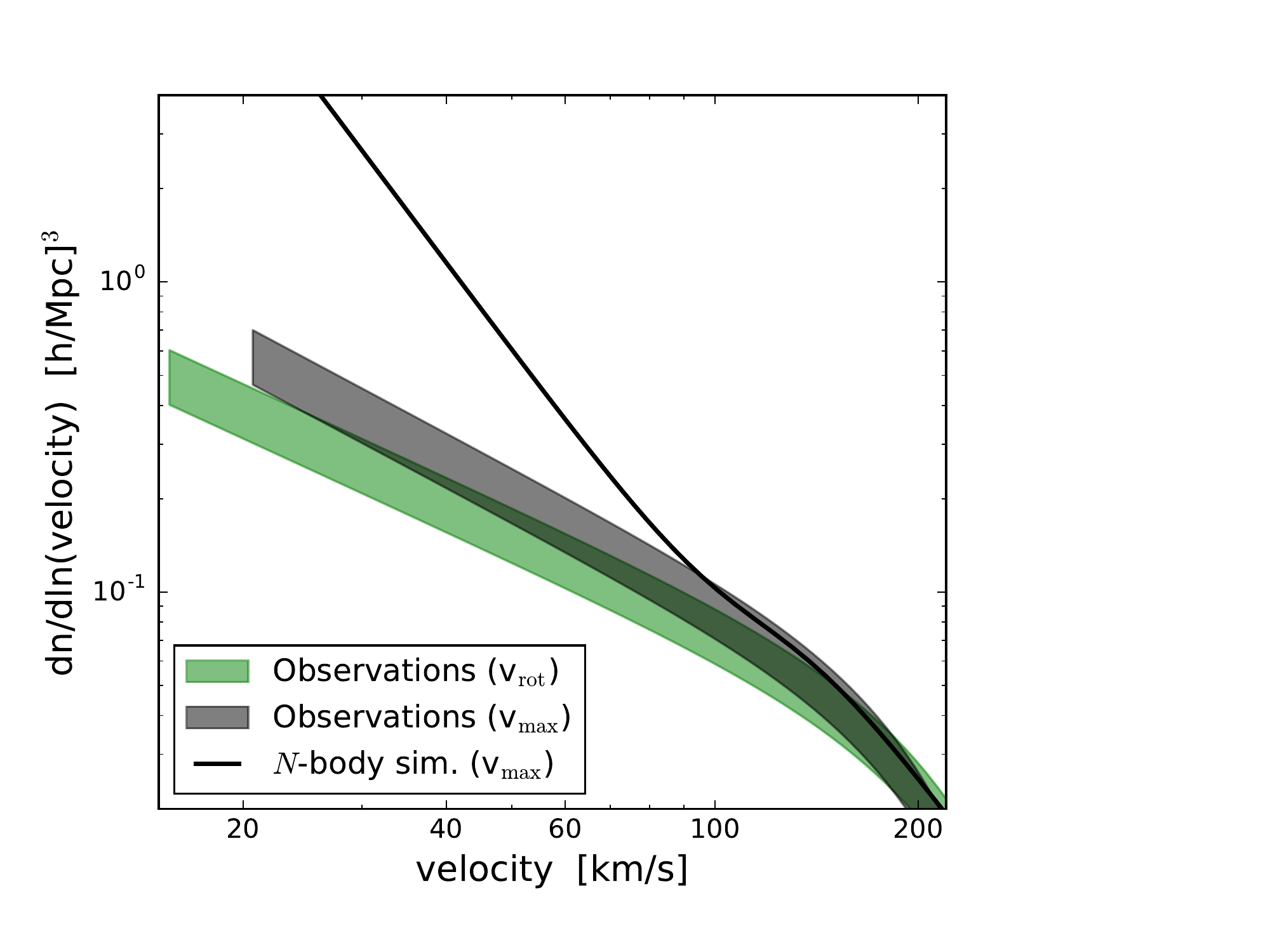}
\caption{\label{fig:intro}\emph{Left:} Observed circular velocities ($v_{\rm out}$ at $r_{\rm out}$) from the selected galaxy sample (black symbols) together with fitted velocity profiles based on NFW (black lines) and NFW+core (red lines, accounting for baryon-induced cores). \emph{Middle:} Relation between $v_{\rm rot}$ of galaxies and $v_{\rm max}$ of haloes from NFW fits (black symbols) and NFW+core fits (red symbols) with error-bars indicating the change of $v_{\rm max}$ when the halo concentrations are varied by 1-$\sigma$ around the mean. Corresponding linear fits are shown as black and red lines. The effects from baryon-induced cores are very small because $r_{\rm out}$ lies well beyond the core radii. \emph{Right:} Velocity function based on $v_{\rm rot}$ \citep[green band, from][]{Klypin:2014ira} and $v_{\rm max}$ (grey band, including $v_{\rm max}$-correction from the middle panel) together with the prediction from gravity-only simulations. The rotation velocity ($v_{\rm rot}$) is defined via HI line width ($w_{50}$) and galaxy inclination ($i$), i.e. $v_{\rm rot}\equiv w_{50}/(2\sin i)$.}}
\end{figure*}

\subsection{Galaxies in the local universe}
In a recent paper, \citet{Klypin:2014ira} performed a detailed analysis of the abundance of galaxies within the local volume around the Milky-Way. Their analysis is based on the galaxy catalogue from \citet[][hereafter K13]{Karachentsev:2013ipr} which they show to be complete down to a limiting magnitude of $M_B=-12$ within 10 Mpc from the Milky-Way. The rotation velocities ($v_{\rm rot}$) of all galaxies in the K13 sample were determined by either relying on inclination corrected unresolved HI line-width measurements or by using the magnitude-velocity relation for the galaxies with no detected HI (a fraction of less than ten percent). \citet{Klypin:2014ira} find a slowly rising VF down to $v_{\rm rot}\sim15$ km/s which they claim to be in tension with the $\Lambda$CDM model prediction below $v_{\rm rot}\sim80$ km/s. Similar conclusions have been made previously by \citet{Zavala:2009ms}, \citet{TrujilloGomez:2010yh}, and \citet{Papastergis:2011xe} based on data from the {\tt HIPASS} and {\tt ALFALFA} HI surveys, respectively\footnote{Recently, \citet{Bekeraite:2016aaa} have shown that there is disagreement between observations and simulations at larger velocities as well (i.e. between 60 km/s and 300 km/s). This tension is, however, not as strong and could be due to the fact that $v_{\rm max}$ of larger galaxies is dominated by the stellar component.}.

The statement that there is tension between observations and the $\Lambda$CDM model relies on the assumption that the rotational velocity from HI line-widths, $v_{\rm rot}\equiv w_{50}/(2\sin i)$, can be used as a proxy for the maximum circular velocity ($v_{\rm max}$) of a halo. \citet{Klypin:2014ira} showed that this is approximately the case, at least for a subset of selected dwarf galaxies and assuming all haloes to have NFW profiles with a given concentration-mass relation. More recently, several authors have questioned the validity of these assumptions, reporting strong biases between $v_{\rm rot}$ and $v_{\rm max}$ instead. These studies used abundance matching \citep{Brook:2015ofa,Brook:2015eva}, measurements from zoom-in hydrodynamical simulations of individual galaxies \citep{Maccio:2016egb}, or semi-analytical models \citep{Yaryura:2016djm,Obreschkow:2013hka}. 

In TG16, we used direct observations of galaxies to perform a detailed investigation of possible biases between $v_{\rm rot}$ and $v_{\rm max}$. We showed that $v_{\rm max}$ can be directly recovered from $v_{\rm rot}$, and we argued that the bias between the two is present but not large enough to solve the discrepancy present in the VF. We will now summarise the method developed in TG16.

\subsection{From $\mathbf{v_{\rm rot}}$ to $\mathbf{v_{\rm max}}$}\label{sec:vmax}
Most of the galaxies from the K13 sample only have HI line-width measurements (or magnitude-based estimates for the subdominant population of gas-free galaxies) without any spatial information, making it impossible to estimate the corresponding $v_{\rm max}$. There is, however, more information for a sub-sample of galaxies with existing spatially resolved measurements of their kinematics. A sample of 200 galaxies with interferometric HI observations ($v_{\rm out})$ at the outermost HI radius ($r_{\rm out}$) was compiled by \citet{Papastergis:2016aba}. We use this catalogue to select all galaxies with $r_{\rm out}>3 r_{1/2}$, where $r_{1/2}$ is the galactic half-light radius (see TG16). This additional selection criterion guarantees that the velocity measurement is not dominated by baryonic effects, including a potential DM core from strong stellar feedback \citep{Read:2015sta}. The final catalogue consists of 109 galaxies (distributed over the full range of relevant scales) which can be used to estimate the relation between $v_{\rm rot}$ and $v_{\rm max}$.

In a first step, we fit NFW profiles \citep{Navarro:1995iw} to the observed velocities $v_{\rm out}$ at radius $r_{\rm out}$ in order to determine the corresponding $v_{\rm max}$, which we then compare to $v_{\rm rot}$ from HI line-width measurements\footnote{All observed values of $v_{\rm out}$ are corrected for pressure support \citep[see Sec.~4.1 in][]{Papastergis:2016aaa}. For large galaxies (with $v_{\rm out}>120$ km/s) we furthermore subtract the expected contribution from stars and cold gas in the galaxy centres. We have checked that this correction does not affect our final results (see TG16 for more details).}. For the concentrations, we use the relation from \citet{Dutton:2014xda} based on {\tt Planck} cosmology. 

In the left panel of Fig.~\ref{fig:intro}, the spatially resolved observed circular velocities $v_{\rm out}$ (black symbols with error bars) are shown together with the fitted velocity profiles based on NFW (grey lines). For most galaxies in the sample, $v_{\rm max}$ is not much larger than $v_{\rm out}$, owing to the fact that $v_{\rm out}$ is observed far out in radius ($r_{\rm out}$). The relation between $v_{\rm rot}$ (i.e. the rotation velocity, obtained via the measurement of unresolved HI line widths) and  $v_{\rm max}$ is shown in the middle panel of Fig.~\ref{fig:intro} (black symbols). The error bars illustrate the sensitivity of the results for varying concentrations within the 1-$\sigma$ scatter given by \citet{Dutton:2014xda}. In general, there is a small bias between $v_{\rm rot}$ and $v_{\rm max}$ slowly growing towards very small velocities.

In a second step, we attempt to include effects from baryons on the circular velocity profiles. We therefore repeat the same analysis using the fit from \citet[][herafter R16]{Read:2015sta} which consists of an NFW profile plus a baryon-induced core proportional to the stellar half-light radius. The mass profile of the R16 fit is given by 
\begin{equation}\label{R16profile}
M_{\rm R16}(r)=M_{\rm nfw}(r)f^n,\hspace{0.3cm}f=\left[\tanh\left(\frac{r}{r_c}\right)\right]
\end{equation}
and is a simple extension of the NFW mass profile ($M_{\rm nfw}$) with two additional free parameters $r_{\rm c}$ and $n$. R16 showed that the sizes of baryon-induced cores are proportional to the half-light radii (i.e. $r_{\rm c}=\eta\, r_{1/2}$) and that $n$ varies between 0 and 1 depending on the individual star formation history of each galaxy. Here we fix $\eta=1.75$ because this was shown by R16 to provide the best match to their simulations. Furthermore, we adopted the value $n=1$ for the second free parameter in order to maximise the effect from baryons.

The effects of the baryon-induced cores are illustrated in the left and middle panel of Fig.~\ref{fig:intro}. The inner part of the velocity profiles are much steeper (left panel, red lines) reflecting how cores affect the velocity profiles. However, the resulting values for $v_{\rm max}$ (middle panel, red symbols) are nearly indistinguishable from the ones obtained with an NFW fit. This shows that baryon-induced cores do not bias the $v_{\rm max}$-estimates of the selected galaxy sample, which is a direct result of our original selection criteria (i.e. $r_{\rm out}>3r_{1/2}$). Such a selection greatly simplifies the analysis and can be justified as long as the resulting galaxy sample is \emph{representative} for the galaxies that make up the VF. In TG16, we used the baryonic Tully-Fisher (BTF) relation to show that this is indeed the case.

\subsection{Corrected velocity function}
Let us now turn our attention to the velocity function and how it can be corrected to account for the bias between the rotational ($v_{\rm rot}$) and the maximum circular velocity ($v_{\rm max}$). This correction is important, since only $v_{\rm max}$ can be directly related to the halo mass (and therefore to the theory prediction) while $v_{\rm rot}$ depends on the details of the gas distribution within the galaxy.

In the right panel of Fig.~\ref{fig:intro} we plot both the observed VF based on $v_{\rm rot}$ (green band) and the predicted VF from gravity-only simulations\footnote{The line is based on the MultiDark suite of simulations \citep{Klypin:2014kpa} and includes a correction for the increase of circular velocities due to the stellar component of galaxies visible beyond 80 km/s \citep[see][]{TrujilloGomez:2010yh}.} of a $\Lambda$CDM universe based on $v_{\rm max}$ (black solid line). The two velocity functions agree reasonably well at large velocities but start to diverge below 100 km/s. This apparent discrepancy between observations and theory has been pointed out repeatedly in the past \citep[see e.g.][]{Tikhonov:2009jq,Zavala:2009ms,TrujilloGomez:2010yh,Papastergis:2011xe}.

It is possible to correct the observed VF using the average relation between $v_{\rm max}$ and $v_{\rm rot}$ obtained with the selected galaxy sample (i.e. the fit from the middle panel of Fig.~\ref{fig:intro}). The resulting $v_{\rm max}$-corrected VF is plotted as a grey band in the right panel of Fig.~\ref{fig:intro}. Despite being slightly steeper, it remains inconsistent with the prediction from gravity-only simulations. In the following, we use this $v_{\rm max}$-corrected VF and compare it to theoretical predictions including baryon effects as well as modifications induced by the DM model.

\section{Theoretical predictions for the velocity function of haloes}\label{sec:theory}
The velocity function (VF) is sensitive to both baryonic feedback effects and the particle nature of dark matter, making an accurate modelling both essential and challenging. In this paper we use an analytical approach based on the extended Press-Schechter (EPS) model. This has the advantage of being easily adaptable to different dark matter models, and it can be used to estimate suppression effects from baryons.

\subsection{Modelling the velocity function}\label{sec:EPS}
The calculation of the halo velocity function is based on the EPS approach presented in \citet{Schneider:2013ria,Schneider:2013wwa}. The first and most important step is to obtain the halo mass function with sharp-$k$ filter
\begin{eqnarray}\label{WDMmassfct}
\frac{dn}{d\ln M}&=&\frac{1}{12\pi^2}\frac{\bar\rho}{M}\nu f(\nu)\frac{P_{\rm lin}(1/R)}{\delta_c^2R^3},\\
\sigma^2(R)&=&\int \frac{d\mathbf{k}^3}{(2\pi)^3} P_{\rm lin}(k)\Theta(1-kR),
\end{eqnarray}
where $P_{\rm lin}(k)$ is the linear power spectrum, $\delta_c=1.686$ the collapse threshold, and $\Theta$ the Heaviside step function. The first crossing distribution $f(\nu)$ is obtained from the ellipsoidal collapse model, yielding
\begin{equation}
f(\nu)=A\sqrt{2\nu/\pi}(1+\nu^{-p}) {\rm e}^{-\nu/2}
\end{equation}
with $\nu= (\delta_c/\sigma)^2$, A=0.322, and p=0.3. The halo mass is assigned to the filter scale with the relation $M=4\pi\bar\rho(cR)^3/3$ where $c=2.5$ \citep[see also][for a similar description]{Benson:2012su}.

In order to obtain the maximum circular velocity, we assume all haloes to be described by an NFW profile. This is a good assumption, even for alternative DM models with extended cores, because the radius corresponding to $v_{\rm max}$ lies beyond the scale radius of the halo. For the concentration-mass relation of the $\Lambda$CDM model, we use the power-law relation
\begin{equation}\label{cDutton}
c(M)=10^{1.025}\left(\frac{10^{12}\, {\rm M_{\odot}/h}}{M}\right)^{0.097}
\end{equation}
from \citet{Dutton:2014xda} based on the {\tt Planck} cosmology. Alternative DM (ADM) scenarios can have different concentrations and we follow the approach from \citet{Schneider:2014rda} which consists of comparing halo formation times between CDM and ADM models and assigning concentrations accordingly. An estimate of the average redshift of halo formation can be obtained by solving the equation
\begin{eqnarray}
D(z_c)&=&\left[1+\sqrt{\frac{\pi}{2}}\frac{1}{\delta_c}\sqrt{\sigma^2(F^{1/3}R)-\sigma^2(R)}\right]^{-1}\label{zcoll}\\
&\equiv& \frac{5\Omega_m}{2}H(z_c)\int_{z_c}^{\infty}dz\frac{(1+z)}{H(z)^3}\nonumber
\end{eqnarray}
for the collapse redshift $z_c(M)$, where $F=0.05$. Once the function $z_c(M)$ is known for both ADM and CDM, we can link together ADM and CDM haloes with the same collapse redshift and assign concentrations for ADM haloes from Eq.~(\ref{cDutton}). Although $z_c(M)$ is a rather poor estimate of the \emph{actual} collapse redshift measured in simulations, the resulting concentrations of ADM haloes are surprisingly accurate \citep[see][]{Schneider:2014rda}.   

To finally obtain the velocity function, we create a mock sample of haloes drawn from the halo mass function, and we assign concentrations from a log-normal distribution. This allows us to determine a value for the maximum circular velocity according to the relation
\begin{equation}
v_{\rm max}=0.465\sqrt{\frac{GM}{r_{\rm vir}}}\left[c^{-1}\ln(1+c)-(1+c)^{-1}\right]^{-1/2}
\end{equation}
\citep[directly resulting from the NFW profile, see][]{Sigad:2000cd} and to re-bin the sample in order to obtain the VF
\begin{equation}
\Phi(v_{\rm max})\equiv\frac{dn}{d\ln v_{\rm max}}.
\end{equation}
Similar approaches have been applied by several authors in the past \citep[see e.g.][]{Zavala:2009ms,Schneider:2013wwa}.

It was shown by \citet{TrujilloGomez:2010yh} and \citet{Dutton:2010dw} that the maximum circular velocity of small galaxies are not affected by baryonic infall or contraction. Larger galaxies with $v_{\rm max}\gtrsim 100$ km/s have boosted velocities due to their baryonic components. We follow \citet{Klypin:2014ira} and correct the maximum circular velocity of massive galaxies by solving the equation
\begin{equation}\label{barcorKlypin}
v_{\rm max}^{\rm dmo}=v_{\rm max}\left[1+\frac{0.35(v_{\rm max}/120\, {\rm kms^{-1}})^6}{1+(v_{\rm max}/120\, {\rm kms^{-1}})^6} \right]^{-1},
\end{equation}
where $v_{\rm max}^{\rm dmo}$ stands for the maximum circular velocity without baryonic correction.

The EPS approach with sharp-$k$ filter has two distinctive advantages with respect to other methods: First of all, it accurately describes the halo abundance of models with arbitrary power spectra, while the standard EPS model with a real-space tophat filter only works for the CDM scenario \citep{Schneider:2014rda}. Second, it does not suffer from artificial clumping, which is a serious problem for direct simulations of DM scenarios with suppressed power spectra \citep[see e.g.][]{Wang:2007he,Lovell:2013ola,Hahn:2015sia,Hobbs:2015dda}.

One drawback of the EPS approach is that it does not account for substructures. To correct for this, we multiply the EPS velocity function by a factor of 1.25 so that it matches the predictions from the MultiDark N-body simulations \citep{Klypin:2014kpa}. This corresponds to adding a constant number of sub-haloes to each velocity bin\footnote{A constant ratio of subhalo to host-halo numbers is a very good approximation for haloes with $v_{\rm max}<150$ km/s. In \citet{Klypin:2010qw} it was shown that this ratio does not change by more than four percent in the range $v_{\rm max}=30-150$ km/s.}. The normalisation is done once and is not changed for different DM models.

\subsection{Maximising effects from baryons}\label{sec:maxbaryoneffects}
The great majority of work on the velocity function has been based on gravity-only $N$-body simulations in the past \citep[e.g.][]{Gonzalez:1999ek,Zavala:2009ms,Zwaan:2009dz,Papastergis:2011xe,Obreschkow:2013hka,Klypin:2014ira}. There are, however, two distinct effects from baryons which should be be accounted for, since they have the potential to significantly alter the VF at dwarf galaxy scales. The first effect is {\it baryonic depletion} and consists of a reduction of the maximum circular velocity due to the fact that some of the gas is being pushed out of haloes, reducing the total mass and accretion rate of the halo during its formation. The second effect is {\it baryonic suppression}, referring to the fact that feedback can reduce the number of observable galaxies by pushing the luminosity below the sensitivity level of a given survey. The maximum suppression of the VF from both types of baryonic effects were quantified in TG16. Here we summarise these results and show how they can be extended to alternative DM (ADM) scenarios.

The maximum effect from \emph{baryonic depletion} can be obtained by calculating the VF for a cosmology where the entire baryon content is removed. This is achieved by replacing $\sigma_8\rightarrow (1-\Omega_b/\Omega_m)\sigma_8$ as well as $\Omega_m\rightarrow (1-\Omega_b/\Omega_m)\Omega_m$, resulting in a scale-independent decrease of the maximum circular velocities, i.e.
\begin{equation}\label{depl}
v^{\rm depl}_{\rm max}\simeq0.86\, v_{\rm max},
\end{equation}
independent of the DM model (see TG16 for more details). We want to stress that this corresponds to the \emph{maximum} baryonic depletion, likely to overestimate the true effect.

The second type of effect, the \emph{baryonic suppression}, is more difficult to model as it crucially depends on the details of the suppression mechanism. In TG16 we developed a model-independent approach to quantify the maximum possible suppression of dwarf galaxy numbers. Any decrease of the stellar or gaseous content of galaxies leads to a bend in the relation between $v_{\rm max}$ and $M_{\rm bar}$ -- the baryonic Tully-Fisher relation of $v_{\rm max}$. The maximum allowed suppression can therefore be directly constrained by the data without prior knowledge of feedback mechanisms\footnote{The argument that baryonic feedback effects suppress the BTF relation is only true if $v_{\rm max}$ is used for the velocity, as it is a direct measure of the halo potential and does not depend on the extent of observable gas (as is the case for $v_{\rm rot}$ for example).}.

In this paper we describe the $v_{\rm max}$-$M_{\rm bar}$ relation with the function $\mathcal{M}(v_{\rm max})$ which provides a good description of the data points. The suppression induced by potential baryonic processes is furthermore parameterised as
\begin{equation}\label{BTFsuppression}
\mathcal{M}_{\rm supp}(v_{\rm max})=[1+(v_s/v_{\rm max})^{4}]^{-5}\mathcal{M}(v_{\rm max})
\end{equation}
where $v_s$ is a free model parameter. This function leads to a very similar suppression than the one obtained in hydro simulations \citep[see e.g.][]{Sales:2016dmm}. A more general parametrisation is discussed in TG16.

For the fiducial case of CDM, the function $\mathcal{M}(v_{\rm max})$ is given by a linear least-square fit of the data. It can be shown that this is indeed a very good fit to the data (see TG16). Next, we can use the \emph{likelihood ratio analysis}\footnote{The logarithmic likelihood ratio test is based on the measure $D\equiv2\ln(\mathcal{L}_m/\mathcal{L}_0)$, where $\mathcal{L}_m$ is the maximum likelihood and $\mathcal{L}_0$ the likelihood of a constrained model with fixed $v_s$. For a large sample size $D$ is known to be $\chi^2$-distributed and a model with given $v_s$ can therefore be excluded at the confidence level (CL) given by the $p$-value from a $\chi^2$ statistic.} to determine the model $\mathcal{M}_{\rm supp}(v_{\rm max})$ with the largest value of $v_c$ that is still in agreement with the data at 3-$\sigma$ confidence level (CL). This function is defined as the model with maximum allowed baryon suppression. For CDM it is given by the parameter $v_s=23$ km/s.

\begin{figure*}
\center{
\includegraphics[width=.328\textwidth,trim={0.6cm 0.7cm 5.57cm 0.4cm}]{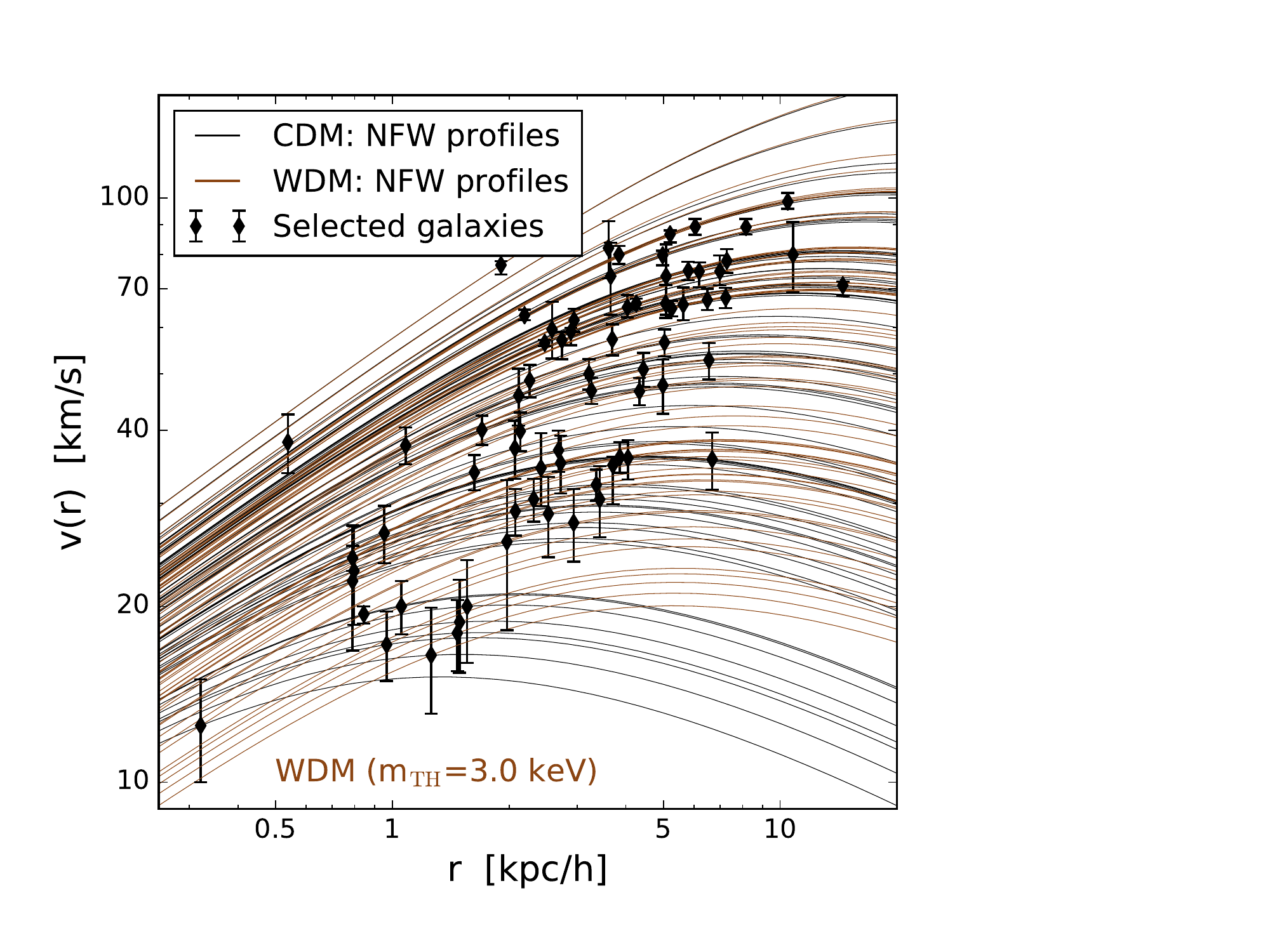}
\includegraphics[width=.328\textwidth,trim={0.6cm 0.7cm 5.57cm 0.4cm}]{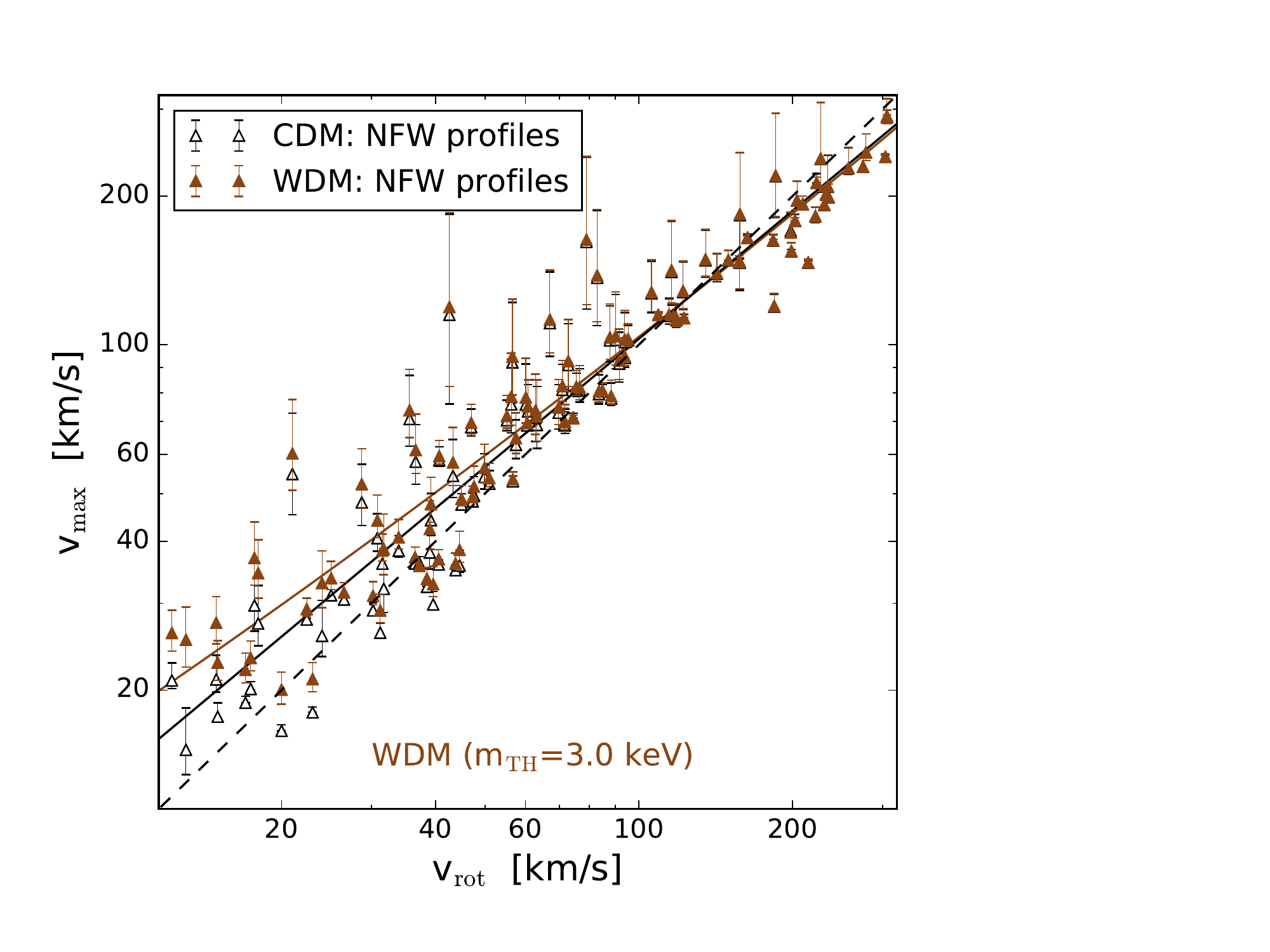}
\includegraphics[width=.328\textwidth,trim={0.6cm 0.7cm 5.57cm 0.4cm}]{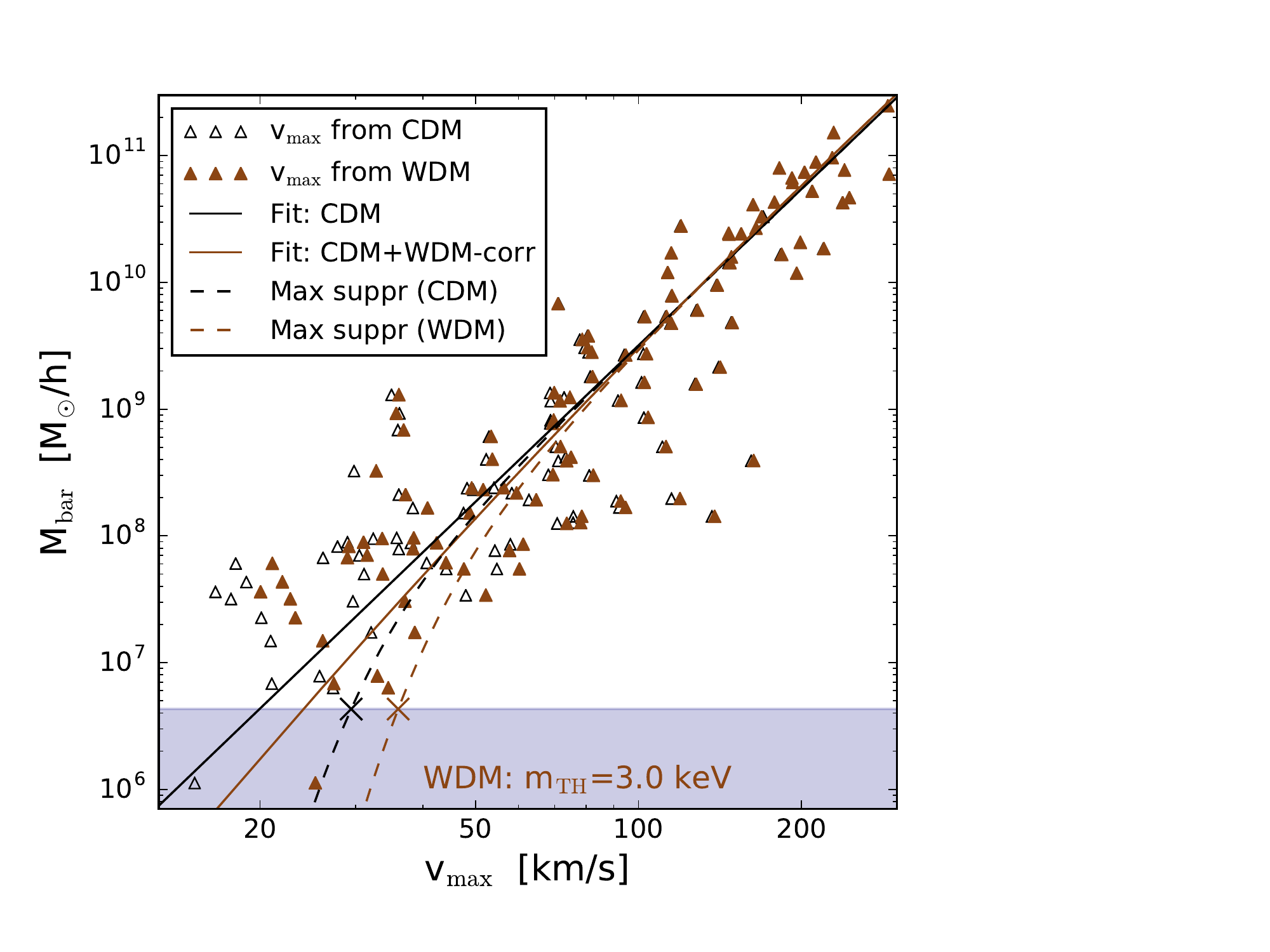}
\caption{\label{fig:WDMprofiles}\emph{Left:} NFW velocity profiles (assuming average concentrations) for CDM (black lines) and WDM (brown lines) fitted to the observed circular velocities of the selected galaxy sample (black symbols with error-bars). \emph{Middle:} Relation between $v_{\rm rot}$ and $v_{\rm max}$ based on the fits from the left panel (empty triangle for CDM, full triangles for WDM). The error-bars indicate the change of $v_{\rm max}$ when the halo concentrations are varied by 1-$\sigma$ around the mean. The black and brown lines are linear and quadratic least square fits, respectively. The rotational velocity ($v_{\rm rot}$) is defined via HI line width ($w_{50}$) and galaxy inclination ($i$), i.e. $v_{\rm rot}\equiv w_{50}/(2\sin i)$. \emph{Right:} Baryonic Tully-Fischer (BTF) relation based on $v_{\rm max}$ from CDM and WDM (empty and full triangles). The solid black line is a linear fit to the BTF relation for CDM. The brown line is obtained from the black line by replacing the mean values of $v_{\rm max}$ from CDM with the ones from WDM (see lines in middle panel). The dashed lines show the strongest allowed suppression of the BTF (3-$\sigma$ excluded by the data), while the crosses indicate the velocities where these lines cross the K13 survey limit (grey band). The zero baryon suppression (solid lines) and the maximum baryon suppression (dashed lines) represent the allowable range of galaxy formation models that we will later use in calculating the VF.}}
\end{figure*}

In the right panel of Fig.~\ref{fig:WDMprofiles} the $v_{\rm max}$-$M_{\rm bar}$ relation of CDM (empty triangles) is plotted together with the linear fit $\mathcal{M}$ (solid black line) and the function of maximum allowed baryon suppression $\mathcal{M}_{\rm supp}$ (dashed black line). While the former is a good fit to the data, the latter is characterised by a strong downturn towards small velocities.

For the alternative DM models discussed in this paper, the values of $v_{\rm max}$ are modified with respect to CDM. As a consequence, the $v_{\rm max}$-$M_{\rm bar}$ relation cannot be described by a linear least-square fit anymore. A more accurate model for $\mathcal{M}$ is obtained when the linear fit from CDM is corrected by accounting for the difference in the average value of $v_{\rm max}$ between ADM and CDM. Based on this corrected function $\mathcal{M}$ for ADM, the maximum baryon suppression model can again be obtained with a likelihood ratio analysis. 

Any downturn in the $v_{\rm max}$-$M_{\rm bar}$ relation is expected to have an influence on the velocity function, as it sets the velocity scale below which galaxies become undetectable by a given survey. At the velocity scale $v_c$, where $\mathcal{M}_{\rm supp}(v_{\rm max})$ crosses the survey detectability limit (in terms of baryon mass), half of the galaxies are too faint to be visible in the velocity function. This effect can be modelled as follows
\begin{eqnarray}\label{VFsuppression}
\Phi_{\rm supp}(v_{\rm max}) &=& \mathcal{G}_{\rm supp}(v_{\rm max}) \Phi(v_{\rm max}),\\
\mathcal{G}_{\rm supp}(v_{\rm max})&=&\frac{1}{2}\left[{\rm erf}\left(\frac{\log v_{\rm max}-\log v_c}{\sqrt{2}\log \sigma_c}\right)+1\right],\nonumber
\end{eqnarray}
where we assumed a log-normal distribution of galaxies around the mean (see TG16 for more details). For the K13 sample, the detectability limit is at $\sim4.3\times 10^6$ M$_{\odot}$/h resulting in $v_c\sim29.5$ km/s for CDM (as indicated by the black cross in the right panel of Fig.~\ref{fig:WDMprofiles}). The scatter can be directly measured from the data (ignoring baryon suppression) resulting in $\log\sigma_c\sim0.15$ for most models (including CDM), except for SIDM where the scatter is larger (see Sec.~\ref{sec:SIDM}).

In the following sections, the theoretical model developed here is applied to cold, warm, mixed, and self-interacting DM models. We always present both the VF without baryon effects as well as the VF with maximum baryon suppression and depletion. These two extreme models quantify the current uncertainty of theory predictions due to unknown feedback effects.


\begin{figure}
\center{
\includegraphics[width=.45\textwidth, trim={1.0cm 1.2cm 4.6cm 0.8cm}]{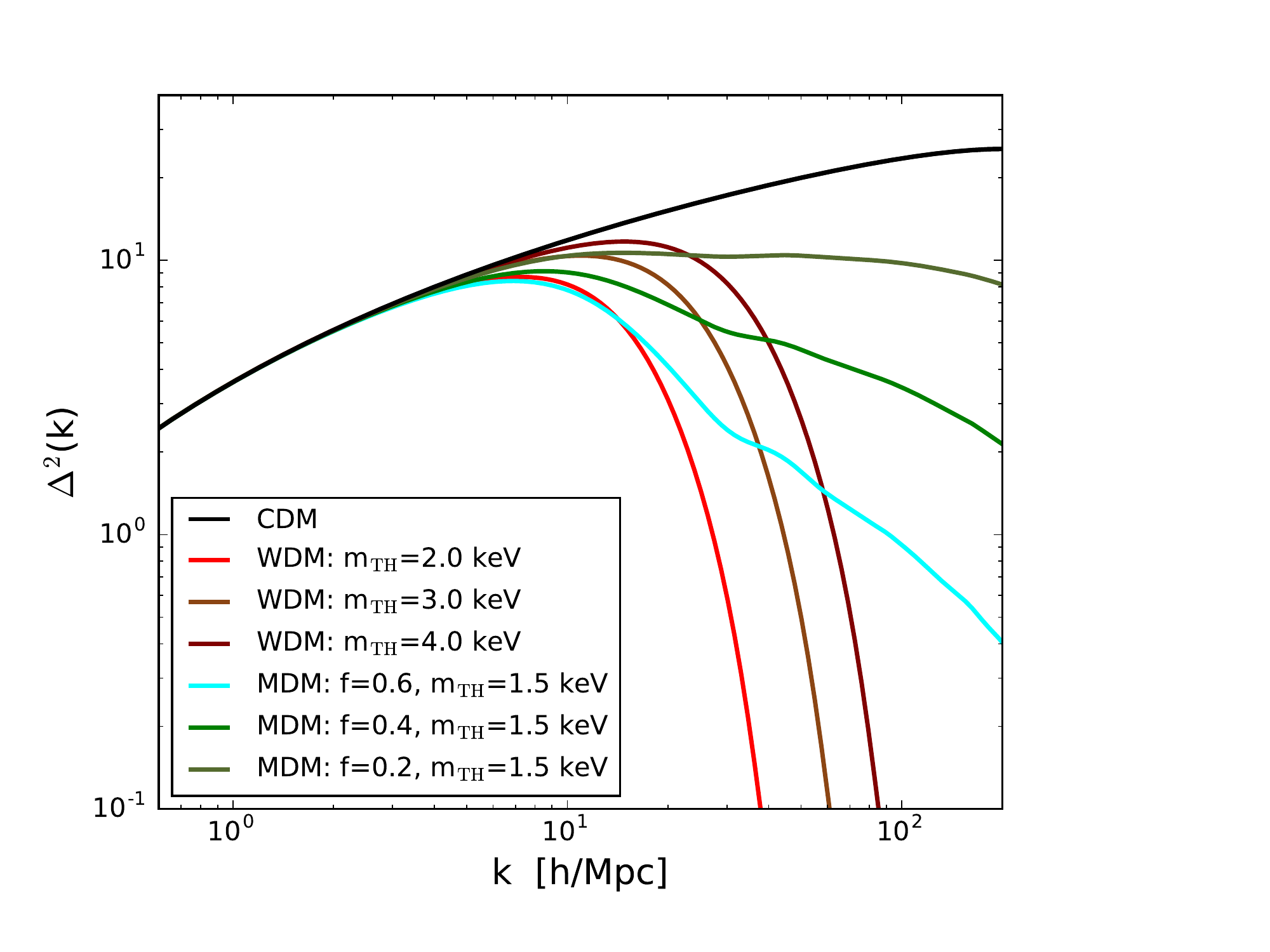}
\caption{\label{fig:powspec}Linear dimensionless power spectra $\Delta^2(k)\equiv k^3P_{\rm lin}(k)/(2\pi^2)$ of the cold, warm, and mixed dark matter models discussed in this paper.}}
\end{figure}

\section{Warm dark matter}\label{sec:WDM}
The first alternative paradigm we consider is the collisionless warm dark matter (WDM) model, which is characterised by a steep suppression of the power spectrum at small scales caused by the free-streaming properties of the DM particles. The scale and exact shape of the suppression depends on both the DM particle mass and its phase-space distribution. In this section, we restrict ourselves to the standard and most studied case of a Fermi-Dirac distributed DM fluid (i.e. the so-called thermal WDM) for which we vary the particle mass. However, other distributions are possible depending on the the exact DM production mechanism \citep[see e.g.][]{Merle:2014xpa}. These models might lead to somewhat shallower suppressions of the power spectrum, closer to the case of mixed DM (see Sec.~\ref{sec:MDM}).

There are various constraints of the thermal WDM scenario from the Lyman-$\alpha$ forest \citep{Seljak:2006qw,Viel:2005qj,Viel:2013apy,Baur:2015jsy}, the dwarf galaxy abundance in the local volume \citep{Polisensky:2010rw,Kennedy:2013uta,Horiuchi:2013noa}, high-redshift galaxies \citep{Menci:2016eww,Menci:2016eui}, or stellar ages of Milky-Way satellites \citep{Chau:2016jzi}. As a rule-of-thumb, the current Lyman-$\alpha$ limits are $m_{\rm TH}\gtrsim 3$ keV while all other limits are around $m_{\rm TH}\gtrsim 2$ keV or weaker\footnote{The constraints from Lyman-$\alpha$ rely on assumptions about the temperature-evolution of the intergalactic medium. Using very high-redshift quasars and assuming a power-law dependence for the temperature, \citet{Viel:2013apy} obtained the limit $m_{\rm TH}\gtrsim 3.3$ keV at 95\% CL. This limit is weakened by about 1 keV if the power-law evolution of the temperature is replaced by an abrupt jump in temperature around $z\sim5$ \citep{Viel:2013apy,Garzilli:2015iwa}. An even stronger constraint of $m_{\rm TH}\gtrsim 4.1$ keV (at 95\% CL) has been found by \citet{Baur:2015jsy} using high-redshift quasars from the {\tt BOSS} survey. This tight limit is, however, relaxed to $m_{\rm TH}\gtrsim 3.0$ keV if cosmological parameters from {\tt Planck} are assumed (instead of the internal parameters from {\tt BOSS}).}.

In this paper we study the representative cases of $m_{\rm TH}=2$, 3, and 4 keV, where the first is in tension with the Lyman-$\alpha$ data but consistent with other limits and the latter two are, roughly speaking, in agreement with observations. The linear power spectra of these models are plotted in Fig.~\ref{fig:powspec} for illustration. They are indistinguishable from CDM at large scales (low wavenumber $k$) but become strongly suppressed towards smaller scales (high $k$). The suppression scale only depends on the thermal mass ($m_{\rm TH}$) of the WDM model.

\begin{figure*}
\center{
\includegraphics[width=.495\textwidth,trim={0.5cm 0.8cm 4.7cm 0.3cm}]{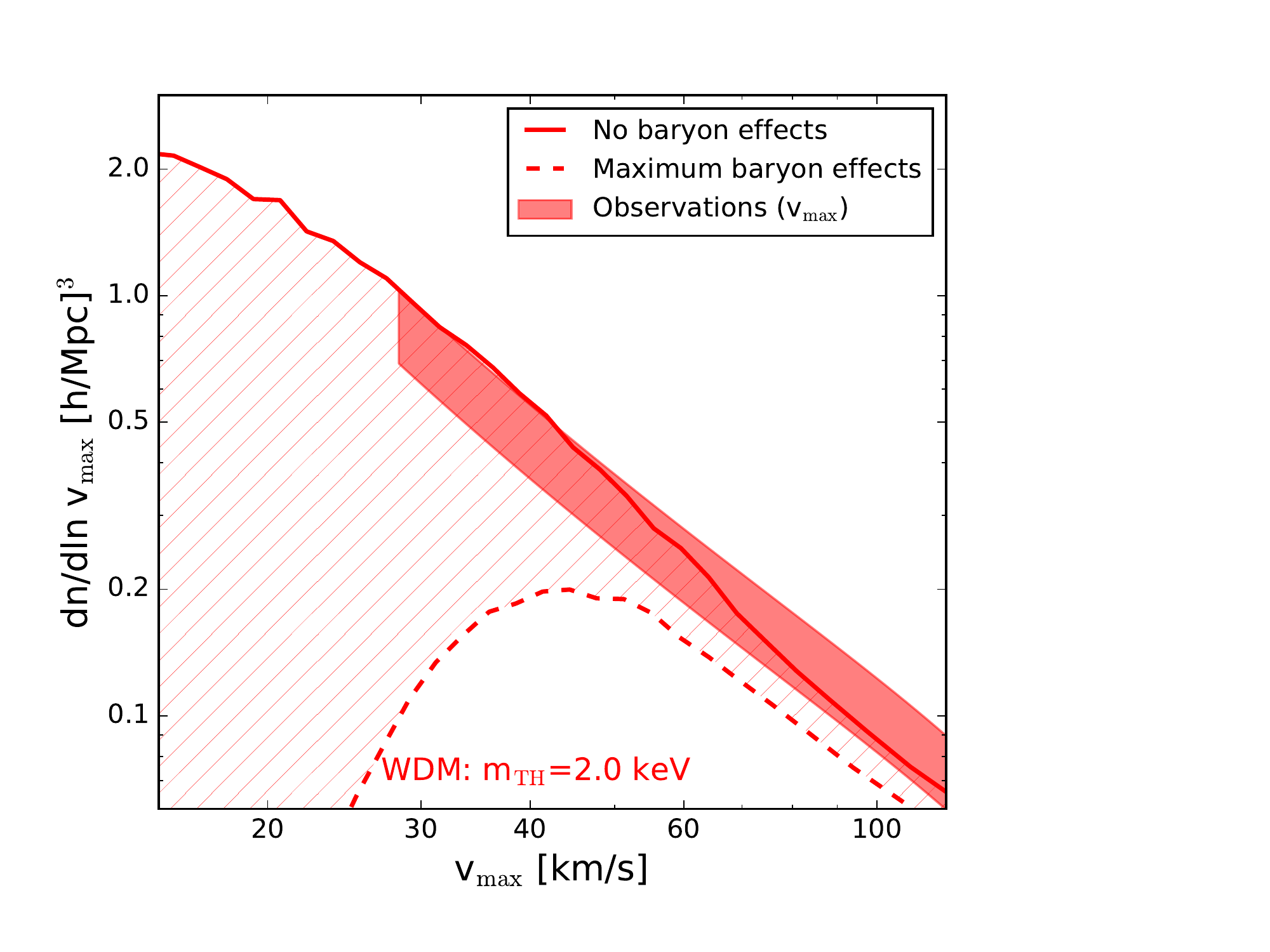}
\includegraphics[width=.495\textwidth,trim={0.5cm 0.8cm 4.7cm 0.3cm}]{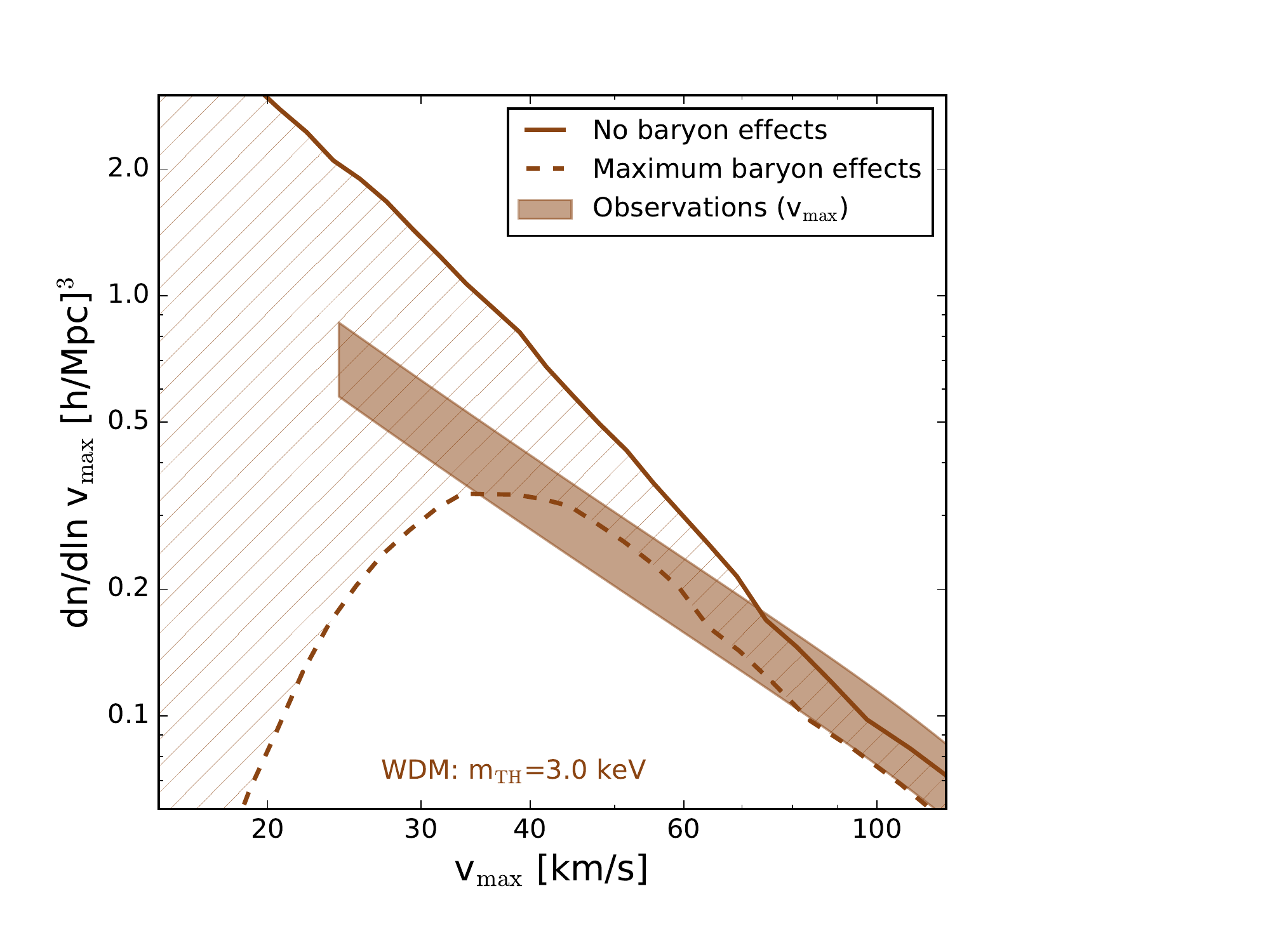}\\
\includegraphics[width=.495\textwidth,trim={0.5cm 0.8cm 4.7cm 0.3cm}]{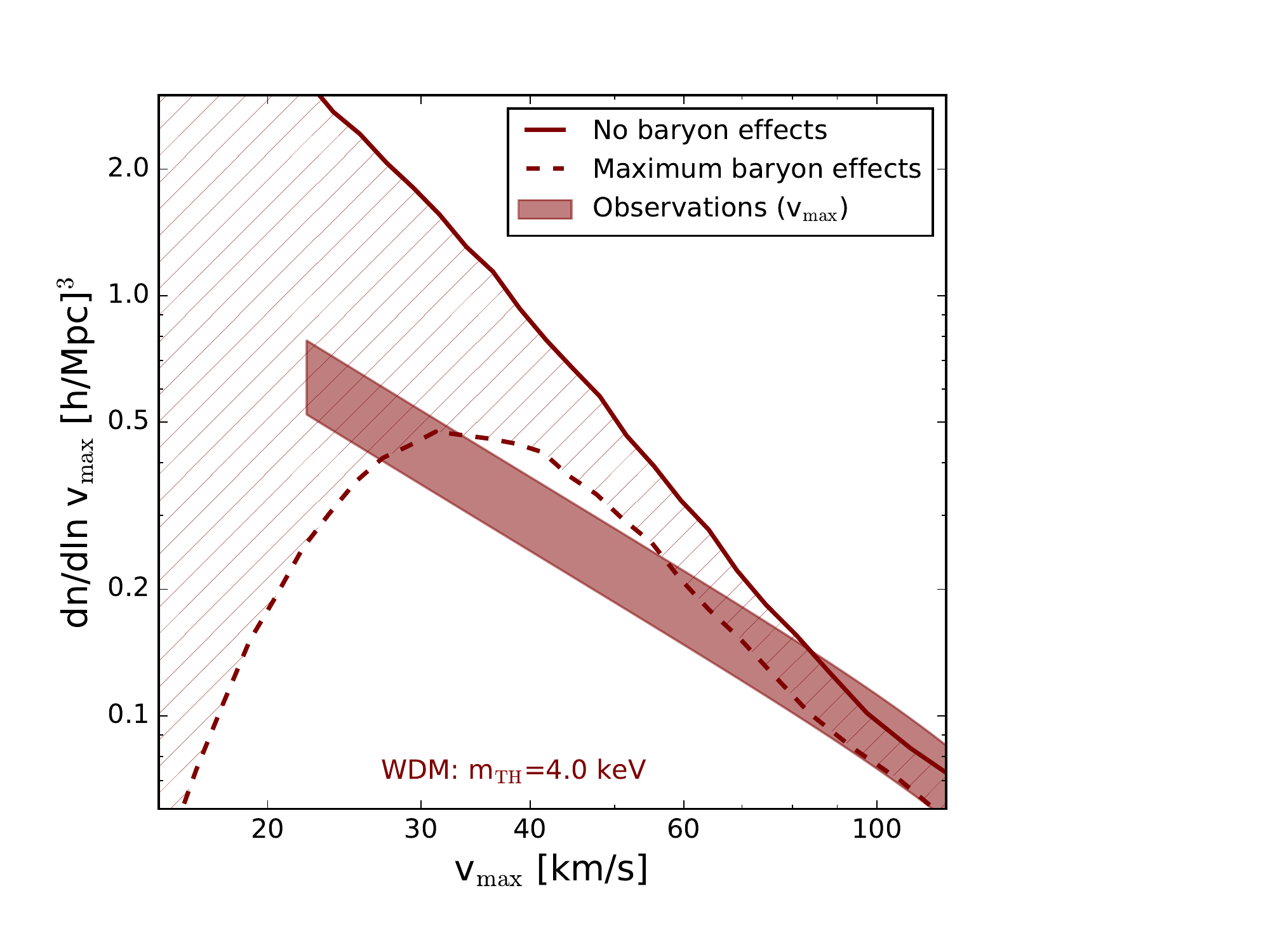}
\includegraphics[width=.495\textwidth,trim={0.5cm 0.8cm 4.7cm 0.3cm}]{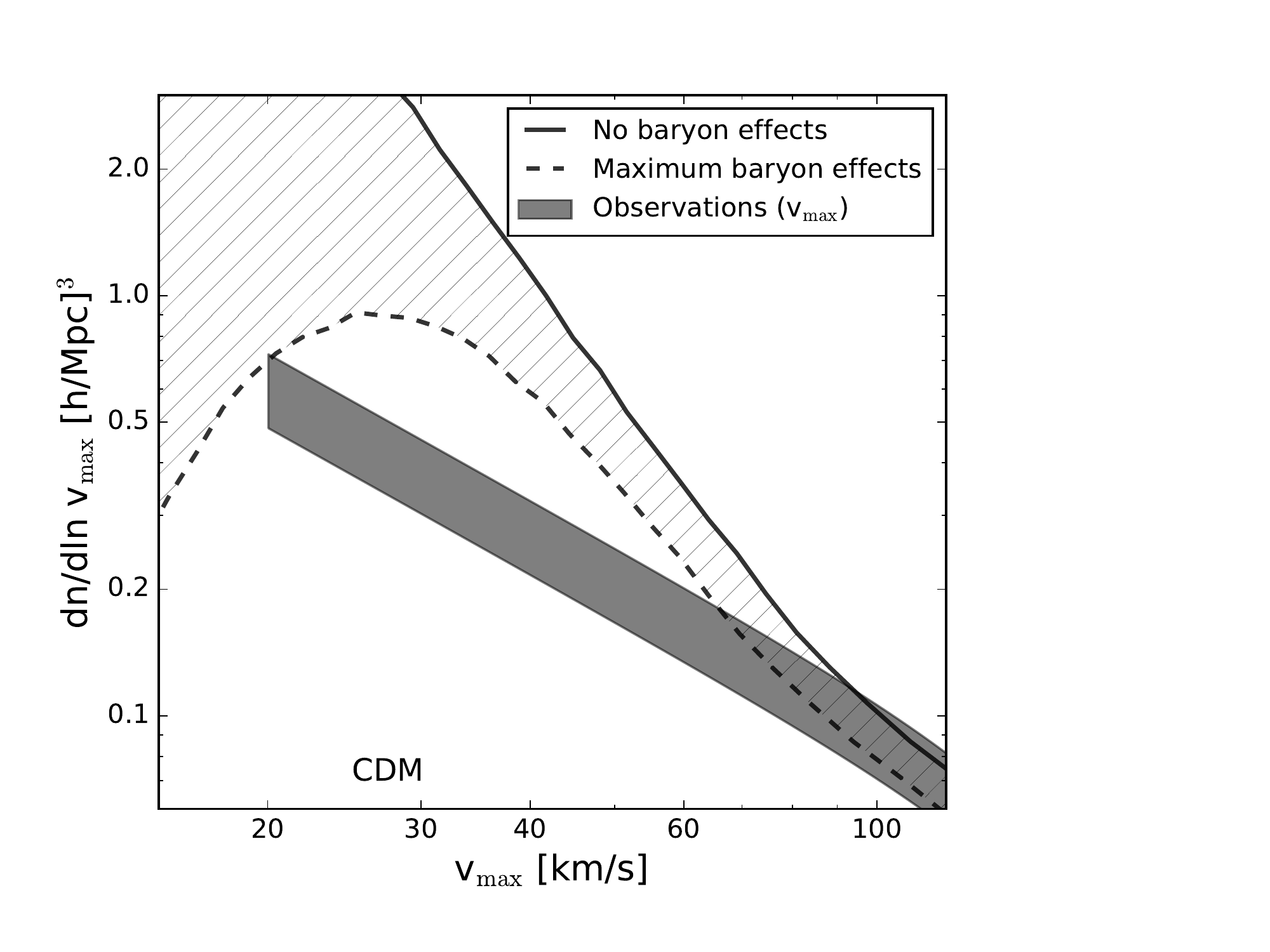}
\caption{\label{fig:WDMvelfct}Velocity functions for (thermal relic) WDM models with $m_{\rm TH}=2$ keV (top left), $m_{\rm TH}=3$ keV (top right), and $m_{\rm TH}=4$ keV (bottom left). The CDM model is shown on the bottom right for comparison. The hatched area between solid and dashed lines illustrates the theoretical uncertainty related to unknown baryon processes (see Sec.~\ref{sec:maxbaryoneffects}). The agreement between observations (shaded band, corrected for $v_{\rm max}$) and theory is significantly better for WDM than for CDM. Given the uncertainties, WDM models between $m_{\rm TH}\sim1.8-3.5$ keV are able to match the observed abundance of galaxies.}}
\end{figure*}

\subsection{Halo profiles}\label{sec:haloprofilesWDM}
It has been shown in the past that WDM models with realistic particle masses do not produce halo cores large enough to be observable \citep{deNaray:2009xj,VillaescusaNavarro:2010qy,Maccio:2012qf,Shao:2012cg}. Instead, the haloes are well described by NFW profiles with a modified concentration-mass relation \citep{Schneider:2014rda,Bose:2015mga,Ludlow:2016ifl}. Rather than monotonically rising towards small masses, as is the case for CDM, the WDM concentrations turn over and decrease again with a maximum around dwarf galaxy scales, with the exact position depending on the DM particle mass \citep{Eke:2000av,Schneider:2011yu}.

The modified concentrations of the WDM model affect the calculation of the maximum circular velocities. In the left panel of Fig.~\ref{fig:WDMprofiles} we show the WDM (with $m_{\rm TH}=3$ keV) circular velocity profiles (brown lines) fitted to the data-points of the outermost rotation measurement from the selected sample galaxies (black symbols). The lower concentrations at small scales lead to larger values of $v_{\rm max}$ further out in radius compared to the case of CDM (black lines). This means that in WDM models, a galaxy with a given $v_{\rm rot}$ can be fit into a more massive DM halo compared to CDM.

In the middle panel of Fig.~\ref{fig:WDMprofiles} we illustrate the difference between WDM and CDM in terms of the relation between $v_{\rm max}$ and $v_{\rm rot}$ (brown and black symbols), where the error-bars indicate the sensitivity of $v_{\rm max}$ to variations in concentration\footnote{The sizes of the error-bars show the maximum variation in $v_{\rm max}$ if the concentrations are raised or lowered by 1-$\sigma$ with respect to the mean value. They do not include observational uncertainties.}. For WDM the data is better fitted by a quadratic fit (brown line) compared to a linear fit for CDM (black line).

The right panel of Fig.~\ref{fig:WDMprofiles} shows the $v_{\rm max}$-$M_{\rm bar}$ relation of the selected galaxy sample for both WDM and CDM (full and empty triangles). While the CDM relation is well described by a linear fit (black solid line), the data bends downwards at small velocities for WDM. We capture this downturn by applying the average shift of $v_{\rm max}$ between WDM and CDM (i.e. the vertical separation between black and brown lines in the middle panel) to the linear fit from CDM, yielding $\mathcal{M}(v_{\rm max})$ for WDM.

Next, we perform a likelihood ratio analysis to find the model $\mathcal{M}_{\rm supp}(v_{\rm max})$ with the maximum baryon suppression. As for CDM, this corresponds to the model with the largest value of $v_s$ allowed by the data at the 3-$\sigma$ CL. This model is shown as dashed brown line in Fig.~\ref{fig:WDMprofiles}. The brown cross indicates the critical velocity ($v_c$) where the line of maximal suppression crosses the completeness limit of the K13 sample. Below this scale, the observed abundance of galaxies could be reduced due to baryonic processes (see Eq.~\ref{VFsuppression}).

In Fig.~\ref{fig:WDMprofiles} we only illustrate the case of WDM with $m_{\rm TH}=3$ keV for brevity. Note, however, that other WDM models show very similar trends with increasing discrepancies between WDM and CDM for decreasing thermal-relic mass $m_{\rm TH}$.

\begin{figure*}
\center{
\includegraphics[width=.328\textwidth,trim={0.6cm 0.7cm 5.57cm 0.4cm}]{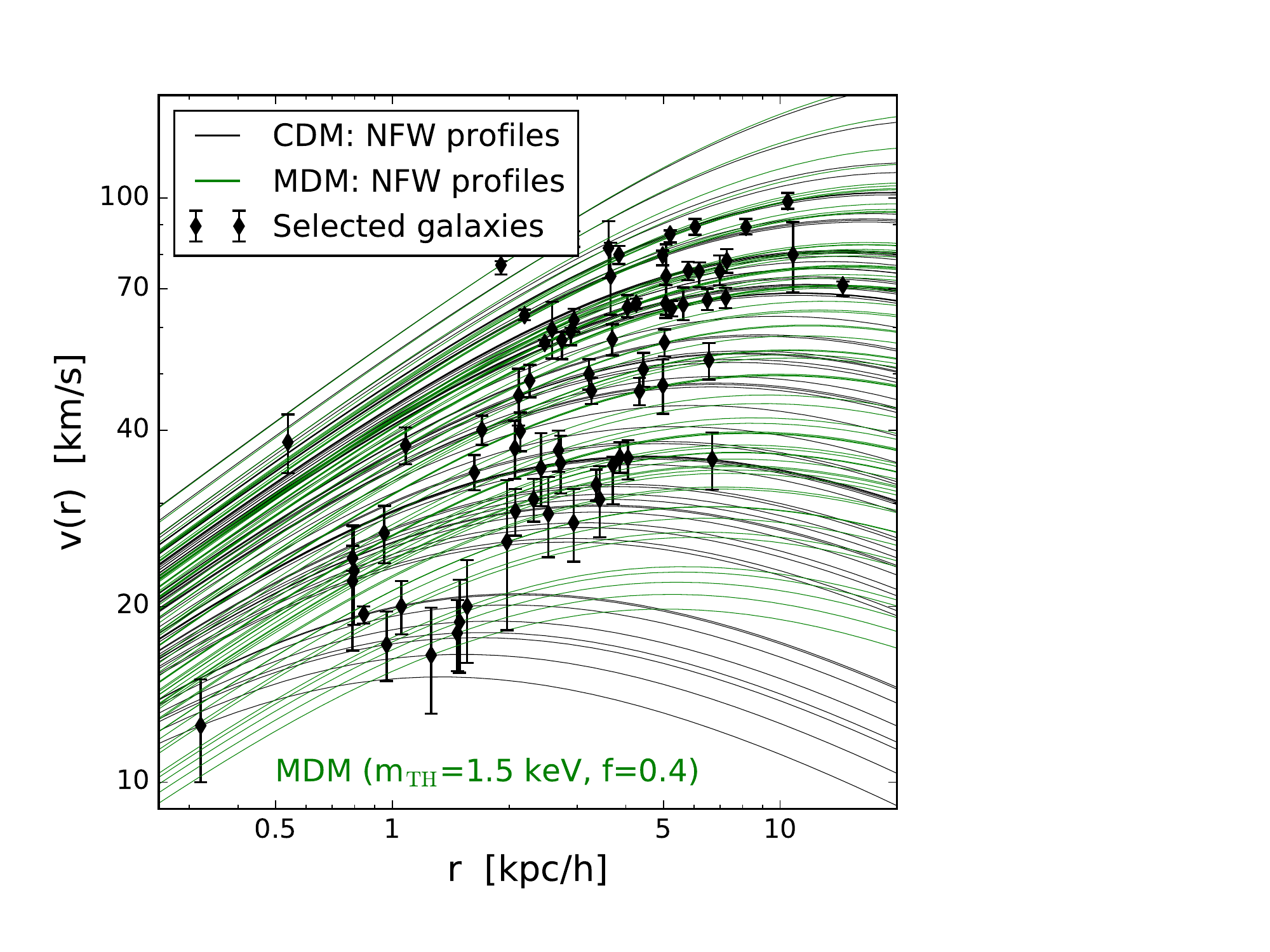}
\includegraphics[width=.328\textwidth,trim={0.6cm 0.7cm 5.57cm 0.4cm}]{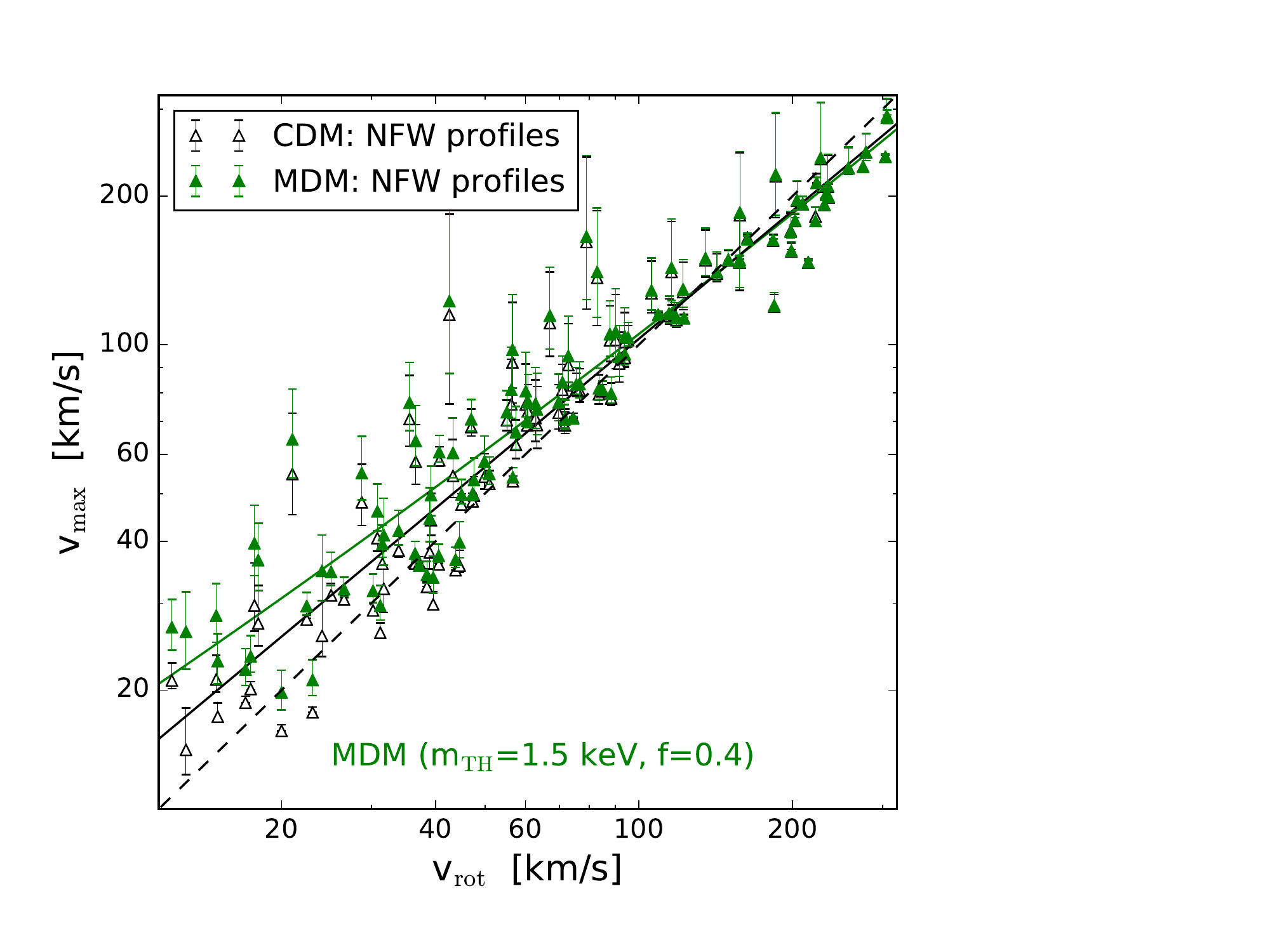}
\includegraphics[width=.328\textwidth,trim={0.6cm 0.7cm 5.57cm 0.4cm}]{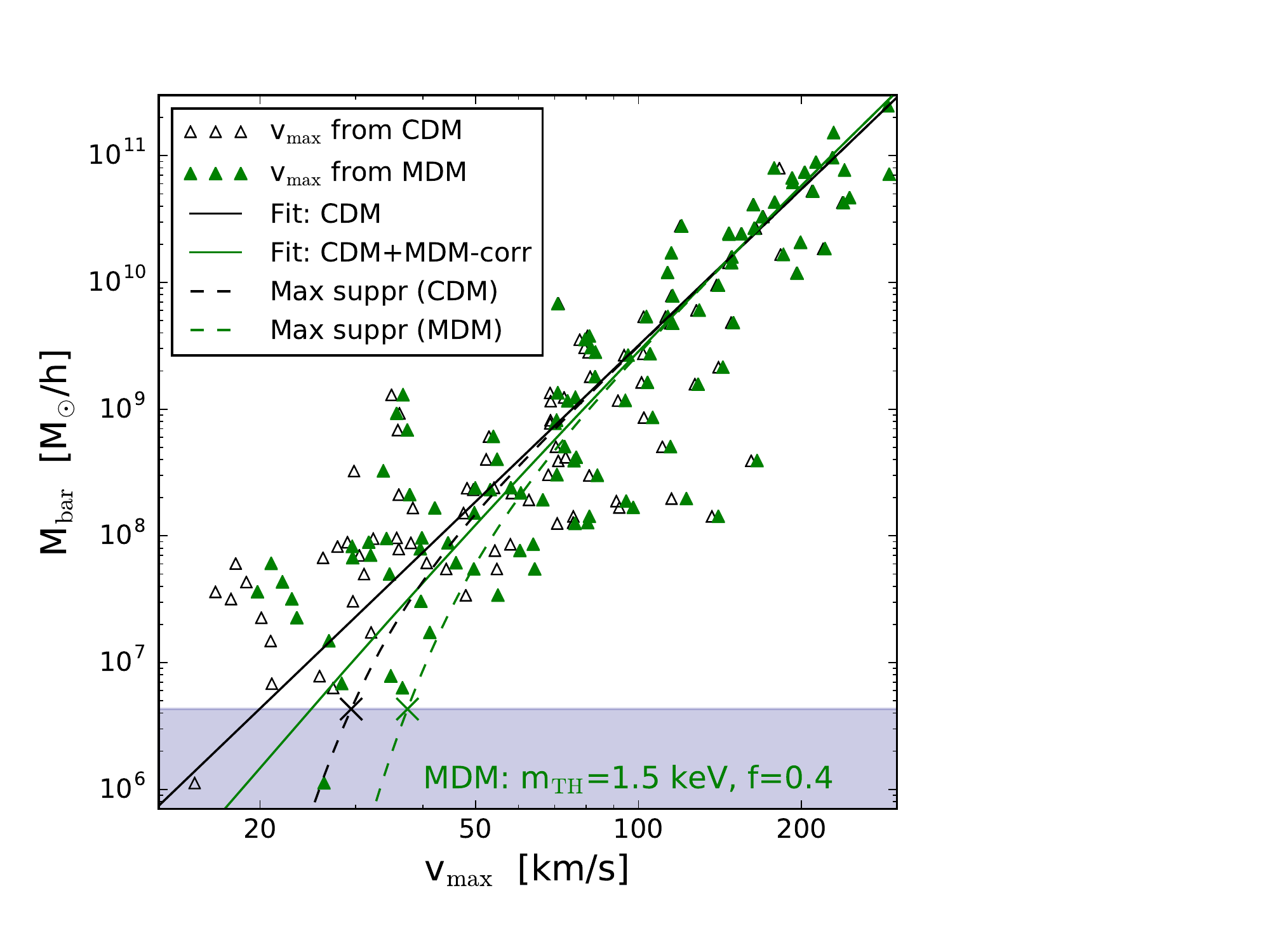}
\caption{\label{fig:MDMprofiles}\emph{Left:} NFW velocity profiles (assuming average concentrations) for CDM (black lines) and MDM (green lines) fitted to the observed circular velocities of the selected galaxy sample (black symbols with error-bars). \emph{Middle:} Relation between $v_{\rm max}$ and $v_{\rm rot}$ based on the fits from the left panel (empty triangle for CDM, full triangles for MDM). The error-bars indicate the change of $v_{\rm max}$ when the halo concentrations are varied by 1-$\sigma$ around the mean. The black and green lines are linear and quadratic least square fits, respectively. The rotational velocity ($v_{\rm rot}$) is defined via HI line width ($w_{50}$) and galaxy inclination ($i$), i.e. $v_{\rm rot}\equiv w_{50}/(2\sin i)$. \emph{Right:} Baryonic Tully-Fischer (BTF) relation based on $v_{\rm max}$ from CDM and MDM (empty and full triangles). The solid black line is a linear fit to the BTF relation for CDM. The green line is obtained from the black line by replacing the mean values of $v_{\rm max}$ from CDM with the ones from MDM (see lines in middle panel). The dashed lines show the strongest allowed suppression of the BTF (3-$\sigma$ excluded by the data), while the crosses indicate the velocities where these lines cross the K13 survey limit (grey band). The zero baryon suppression (solid lines) and the maximum baryon suppression (dashed lines) represent the allowable range of galaxy formation models that we will later use in calculating the VF.}}
\end{figure*}

\subsection{Velocity function}
Assuming a WDM scenario affects the velocity function (VF) in several non-trivial ways. First of all, the predicted VF is flatter in WDM than in CDM  due to a combination of lower halo abundance and lower concentrations. The former reduces the number of observable galaxies while the latter lowers the maximum circular velocity at a given mass scale. Second, the $v_{\rm max}$-corrected VF from observations becomes steeper in WDM compared to CDM, the reason being the modified relation between $v_{\rm rot}$ and $v_{\rm max}$. This is again a direct consequence of the reduced concentrations which cause galaxies of a given $v_{\rm rot}$ to inhabit larger haloes. Both effects are expected to improve the agreement between theory and observations for WDM compared to CDM. 

In Fig.~\ref{fig:WDMvelfct} we show the velocity function of three different WDM models with thermal masses of $m_{\rm TH}=2$ keV (top left), $m_{\rm TH}=3$ keV (top right), and  $m_{\rm TH}=4$ keV (bottom left) as well as the VF for CDM (bottom right). Both the flattening of the predicted and the steepening of the observed $v_{\rm max}$-VF are well visible in the plot, the effects becoming more pronounced for decreasing values of $m_{\rm TH}$.

In each panel of Fig.~\ref{fig:WDMvelfct}, the predicted VF with no baryon effects and with maximum baryon effects are shown as solid and dashed lines. While \emph{baryon depletion} induces a horizontal shift towards small velocities, \emph{baryon suppression} leads to a turn-over of the VF below a characteristic velocity ($v_{c}$). The value of $v_c$ depends on the model and becomes larger for a smaller thermal relic mass ($m_{\rm TH}$). In general, the area between these lines (hatched area) illustrates the theoretical uncertainty related to unknown baryon effects.

It is obvious from Fig.~\ref{fig:WDMvelfct} that the WDM models lead to a better match between theory and observations than CDM. For the models with $m_{\rm TH}=2$ and 3 keV, the $v_{\rm max}$-VF from observations overlaps with the theory prediction (given the uncertainties in baryonic effects). For the coolest model with $m_{\rm TH}=4$ keV, a small tension between observation and theory starts to be visible around $v_{\rm max}\sim40$ km/s, but the discrepancy is still significantly smaller than for the case of CDM (reproduced in the bottom right panel of Fig.~\ref{fig:WDMvelfct}).

It is remarkable that warm DM models with thermal masses of $m\gtrsim3$ keV are able to solve (or at least significantly alleviate) the problem of the VF. These models are cold enough to agree with the very stringent bounds from the Lyman-$\alpha$ forest, therefore offering a truly viable alternative to standard CDM. Former studies have argued that only extreme WDM scenarios, which are in conflict with constraints from Lyman-$\alpha$ and MW satellite counts, are able to solve the mismatch of the VF \citep{Zavala:2009ms,Papastergis:2011xe,Schneider:2013wwa,Klypin:2014ira,Papastergis:2014aba}. These studies, however, did not account for suppression effects from the baryon sector nor for the correction of the observed VF due to larger values of $v_{\rm max}$ for WDM.

\subsection{The special case of sterile neutrinos}
The sterile (or right-handed) neutrino is often considered as the prime candidate for WDM. It is a well motivated hypothetical particle based on a straightforward extension of the standard model neutrino sector. Sterile neutrinos can only play the role of DM if their mass is in the keV-range, otherwise they would either not cluster enough or decay too quickly \citep[see e.g.][]{Adhikari:2016bei}.

A popular way to produce sterile neutrino DM in the early universe is via resonant mixing with active neutrinos \citep{Shi:1998km,Abazajian:2001nj,Asaka:2005an}. This production mechanism differs from thermal freeze-out and does not lead to Fermi-Dirac like momentum distributions. As a result, the suppression in the power spectrum can be somewhat shallower than for the case of the standard (thermal relic) WDM, at least for parts of the parameter space \citep{Ghiglieri:2015jua,Venumadhav:2015pla}.

A similar effect is observed if sterile neutrino DM is produced via the decay of a heavy scalar singlets \citep{Kusenko:2006rh,Shaposhnikov:2006xi}. Depending on the coupling of the scalar to the standard model and on the decay width, the resulting sterile neutrino momentum distribution can strongly differ from a Fermi-Dirac function and may lead to shallower suppressions of the power spectra \citep{Merle:2015oja,Merle:2015vzu,Konig:2016dzg}.  

In terms of the velocity function, sterile neutrino DM is expected to show qualitatively similar effects to the thermal-like WDM models \citep[see e.g.][]{Lovell:2016fec}. At the quantitative level, some differences are expected due to changes in the shape of the power spectra \citep{Schneider:2016uqi}. A detailed investigation of the effects of sterile neutrino DM on the VF will be performed in future work. 


\begin{figure*}
\center{
\includegraphics[width=.495\textwidth,trim={0.5cm 0.8cm 4.7cm 0.3cm}]{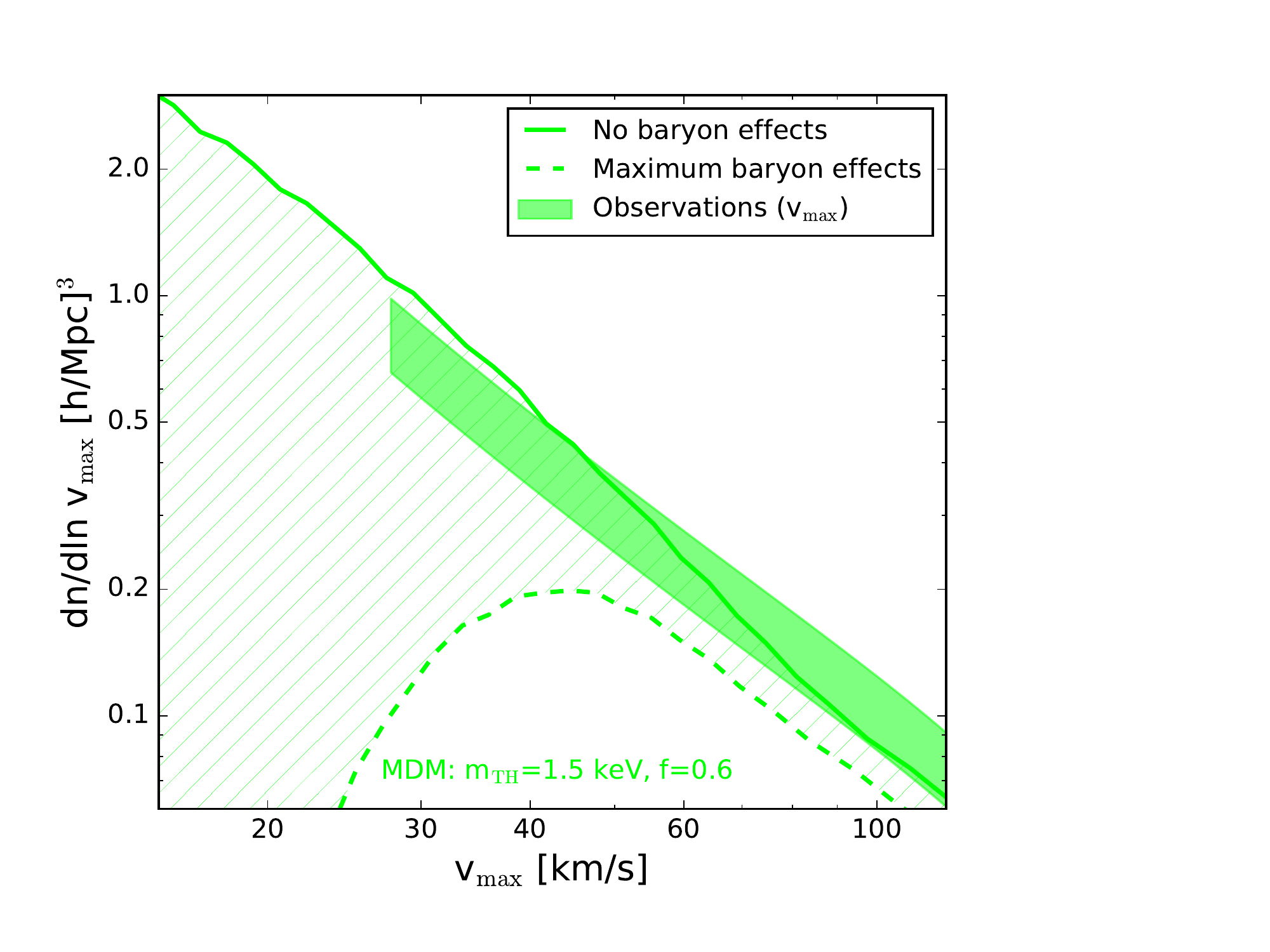}
\includegraphics[width=.495\textwidth,trim={0.5cm 0.8cm 4.7cm 0.3cm}]{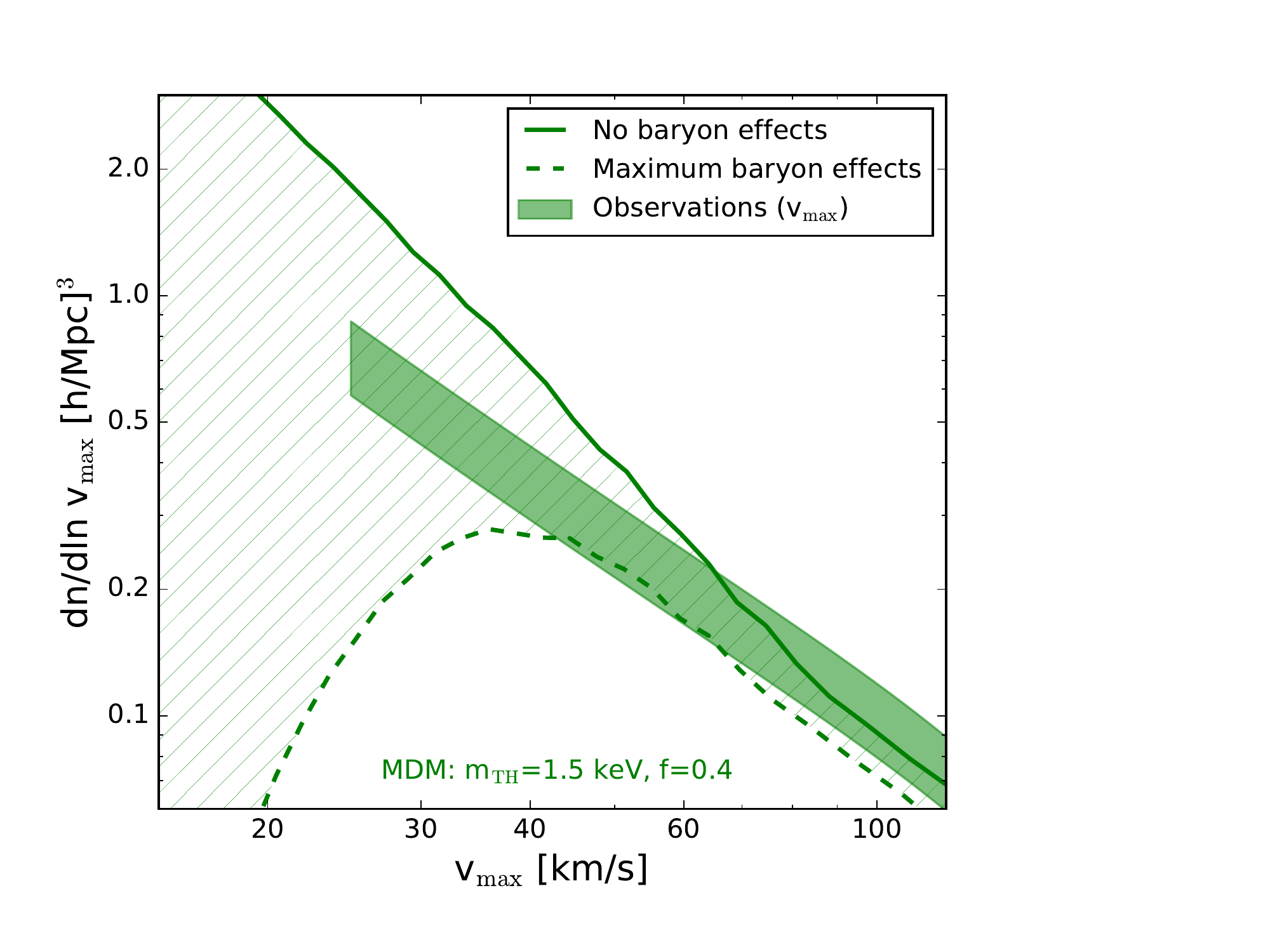}\\
\includegraphics[width=.495\textwidth,trim={0.5cm 0.8cm 4.7cm 0.3cm}]{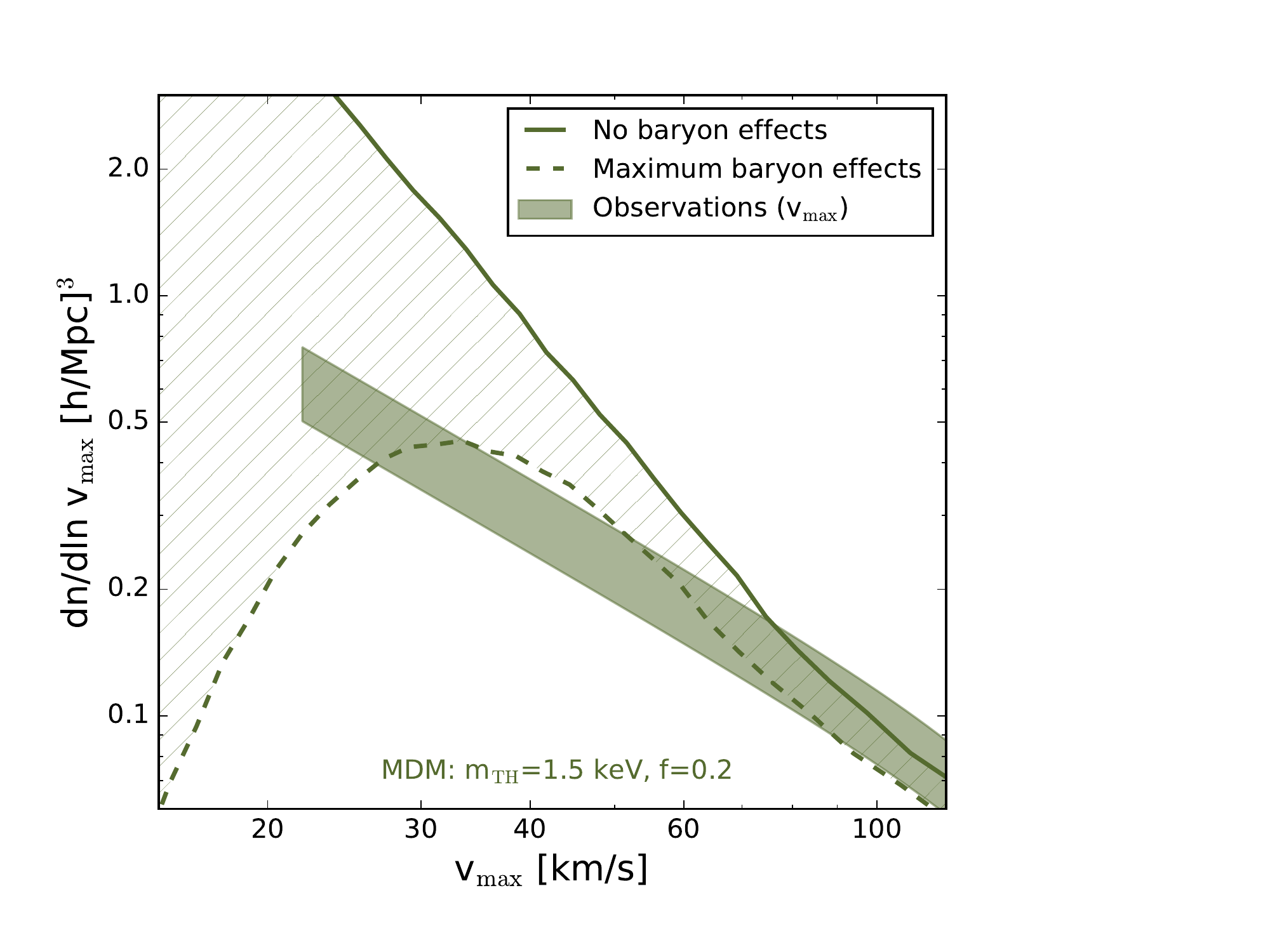}
\includegraphics[width=.495\textwidth,trim={0.5cm 0.8cm 4.7cm 0.3cm}]{Figs/velfct_CDM.pdf}
\caption{\label{fig:MDMvelfct}Velocity functions for MDM models (i.e. a mixture of WDM and CDM species) with $m_{\rm TH}=1.5$ keV and a varying WDM fraction of $f=0.6$ (top left), $f=0.4$ (top right), and $f=0.2$ (bottom left). The CDM model is shown on the bottom right for comparison. The hatched area between solid and dashed lines illustrates the theoretical uncertainty related to unknown baryon processes (see Sec.~\ref{sec:maxbaryoneffects}). The agreement between observations (shaded band, corrected for $v_{\rm max}$) and theory is significantly better for MDM than for CDM.}}
\end{figure*}

\section{Mixed dark matter}\label{sec:MDM}
A straight forward extension to the dark sector is to assume more than one DM species. There is a wealth of possibilities for MDM scenarios including different particle species with different kinds of couplings. These range from two unrelated purely gravitationally interacting DM species to phenomenologically rich scenarios which mirror the baryonic sector. In fact, one could argue that we are already confronted with a MDM universe, since neutrinos are massive, behave exactly like a dark matter fluid, and have a non-negligible effect on structure formation.

In this section we limit ourselves to the simple case of a mixture between cold and (thermal relic) warm DM \citep[see e.g.][]{Borgani:1996ag,Palazzo:2007gz,Boyarsky:2008xj}. This is a hypothetical model with the advantage of yielding a large variety of suppressed power spectra. Similar to WDM, it reduces the number of small galaxies \citep{Anderhalden:2012jc,Maccio':2012uh} and produces halo profiles with smaller concentrations \citep{Schneider:2014rda}, but the affected mass range can be much larger.

In addition to the particle mass of the warm component ($m_{\rm TH}$), the MDM model is characterised by the mass fraction of warm to cold species, i.e. $f=\Omega_{\rm WDM}/(\Omega_{\rm WDM}+\Omega_{\rm CDM})$. In this paper we investigate three cases with the same WDM particle mass of $m_{\rm TH}=1.5$ keV and different fractions $f=0.2$, 0.4, and 0.6. Of course, this only covers a very small fraction of the full MDM parameter space, but it serves as an illustration for the kind of corrections that can be expected for other combinations of mixed DM.

The linear power spectra of the three MDM models are shown in Fig.~\ref{fig:powspec}. They are suppressed with respect to CDM but the shape of the suppression is much shallower than for WDM, spanning many orders of magnitudes in length scale. 


\subsection{Halo profiles}
The haloes of the mixed dark matter scenario are well described by NFW profiles \citep{Anderhalden:2012qt,Maccio':2012uh} with reduced concentrations at small scales. The shape of the concentration-mass relation can again be directly obtained from the linear power spectrum of MDM by assigning the same concentrations to haloes with the same collapse redshift \citep[see][]{Schneider:2014rda}.

Fig.~\ref{fig:MDMprofiles} shows the velocity profiles (left panel) as well as the $v_{\rm max}$-$v_{\rm rot}$ dependence (middle panel) for a mixed DM model with $f=0.4$ and $m_{\rm TH}=1.5$ keV. Similarly to the case of WDM, the values of $v_{\rm max}$ are increased for small galaxies in MDM with respect to CDM. The resulting relation between $v_{\rm max}$ and $v_{\rm rot}$ is well fitted by a curved line which flattens out towards small velocities. The flattening starts at slightly larger scales than for WDM, owing to the smaller mass of the warm component.

The maximum allowed baryon suppression for mixed dark matter is obtained in the same way as for the WDM scenario (see Sec.~\ref{sec:haloprofilesWDM}). First, we define the function $\mathcal{M}$ describing the $v_{\rm max}$-$M_{\rm bar}$ relation of galaxies in MDM. Then, we determine the maximum allowed baryon suppression using the likelihood ratio analysis. The corresponding function ($\mathcal{M}_{\rm supp}$), is shown as dashed green line in Fig.~\ref{fig:MDMprofiles}. Finally, we use the completeness limit of the K13 sample to determine the characteristic velocity $v_c$ (green cross). The value of $v_c$ is model dependent (increasing with higher fractions $f$) and sets the largest scale at which the velocity function could be affected by baryon suppression.

For brevity, we illustrate only one MDM scenario in Fig.~\ref{fig:MDMprofiles}. However, other models show similar trends with growing discrepancies between MDM and CDM for increasing fraction $f$ or decreasing mass $m_{\rm TH}$.

\subsection{Velocity function}
Within the mixed dark matter scenario, the velocity function is affected in a similar way to the case of WDM. First of all, the predicted VF is shallower than in the case of CDM, owing to a combination of reduced halo abundance and concentrations. Second, the observed and $v_{\rm max}$-corrected VF becomes steeper because of higher estimates of $v_{\rm max}$ in MDM as opposed to CDM.

Both effects are visible in Fig.~\ref{fig:MDMvelfct}, where we plot the MDM models with $m_{\rm TH}=1.5$ keV and $f=0.6$ (top left), $f=0.4$ (top right), and $f=0.2$ (bottom right). Not surprisingly, all three models provide a much better match between theory and observations than in the case of CDM (bottom right). For the first two models, there is full agreement between $(v_{\rm max}$-corrected) observations (shaded bands) and theory predictions that include the uncertainties of baryon effects (hatched areas, bracketed by the solid and dashed lines). A small tension starts to be visible for the model with $f=0.2$, but the discrepancy between theory and observations remains significantly smaller than for the case of CDM.

The example of MDM illustrates that many alternative DM models have the potential to alleviate the problem of the over-abundance of field galaxies, provided they suppress perturbations at small scales. In Sec.~\ref{sec:othercandidates} we will briefly discuss some other models with similar characteristics.


\begin{figure}
\center{
\includegraphics[width=.48\textwidth,trim={0.8cm 1.0cm 4.6cm 1.0cm}]{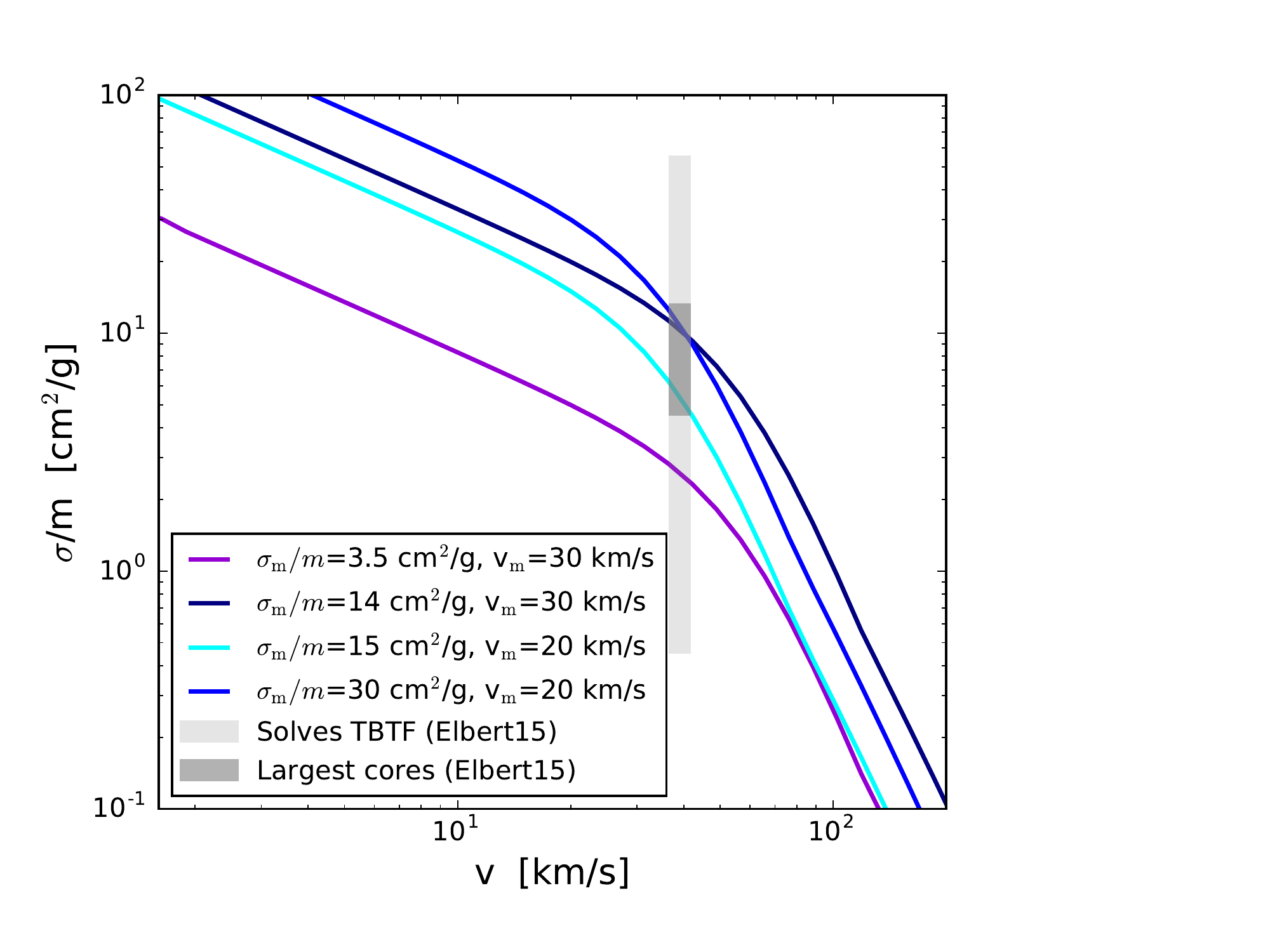}
\caption{\label{fig:crosssections}Velocity-dependent cross sections for the self-interacting dark matter models studied in this paper. The light and dark grey bands indicate the range of scales where the TBTF problem is alleviated and where the largest cores are expected \citep[see][]{Elbert:2014bma}. The purple model corresponds to a scenario investigated in \citet{Vogelsberger:2012ku}}}
\end{figure}

\begin{figure*}
\center{
\includegraphics[width=.328\textwidth,trim={0.6cm 0.7cm 5.57cm 0.4cm}]{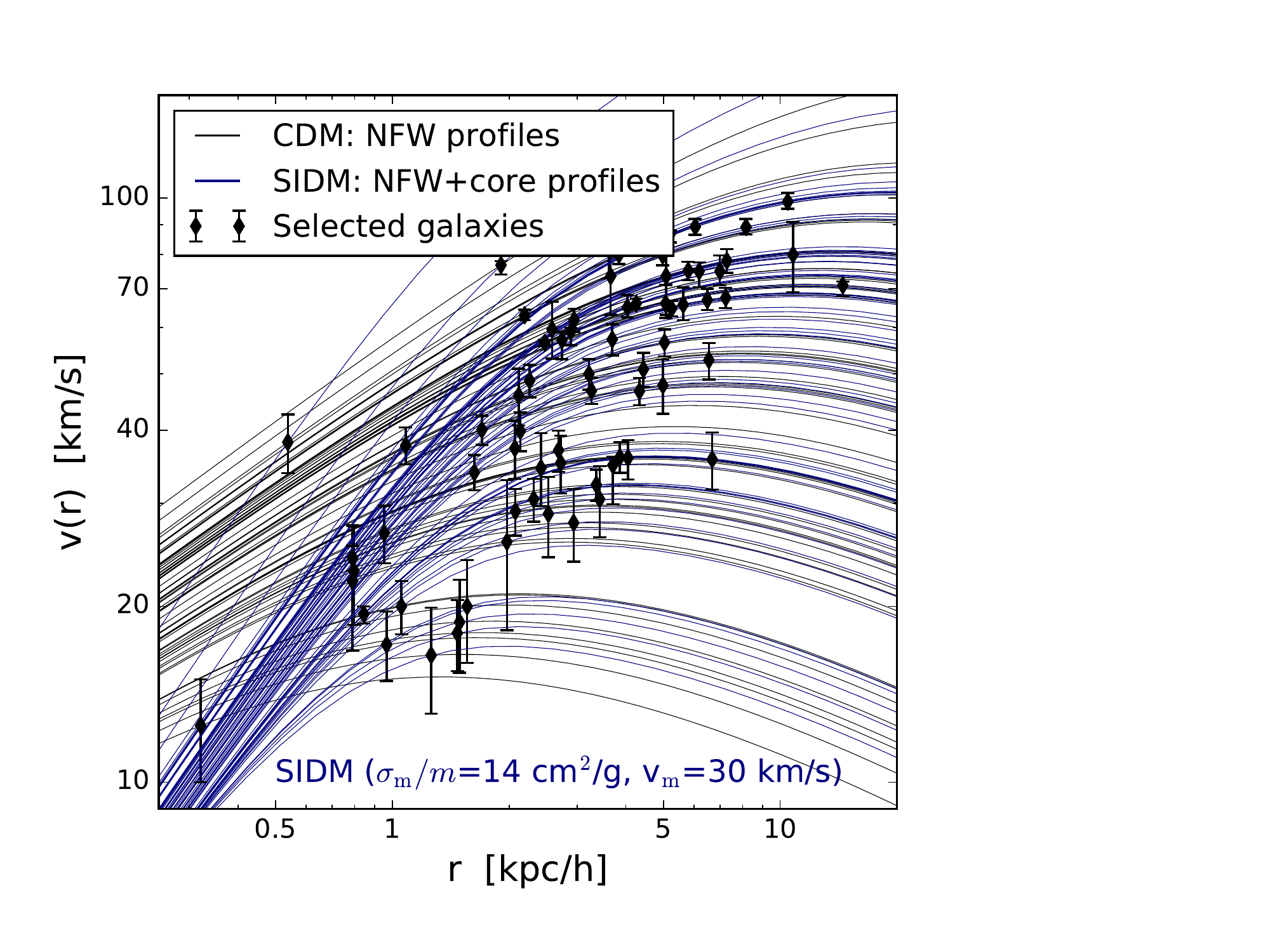}
\includegraphics[width=.328\textwidth,trim={0.6cm 0.7cm 5.57cm 0.4cm}]{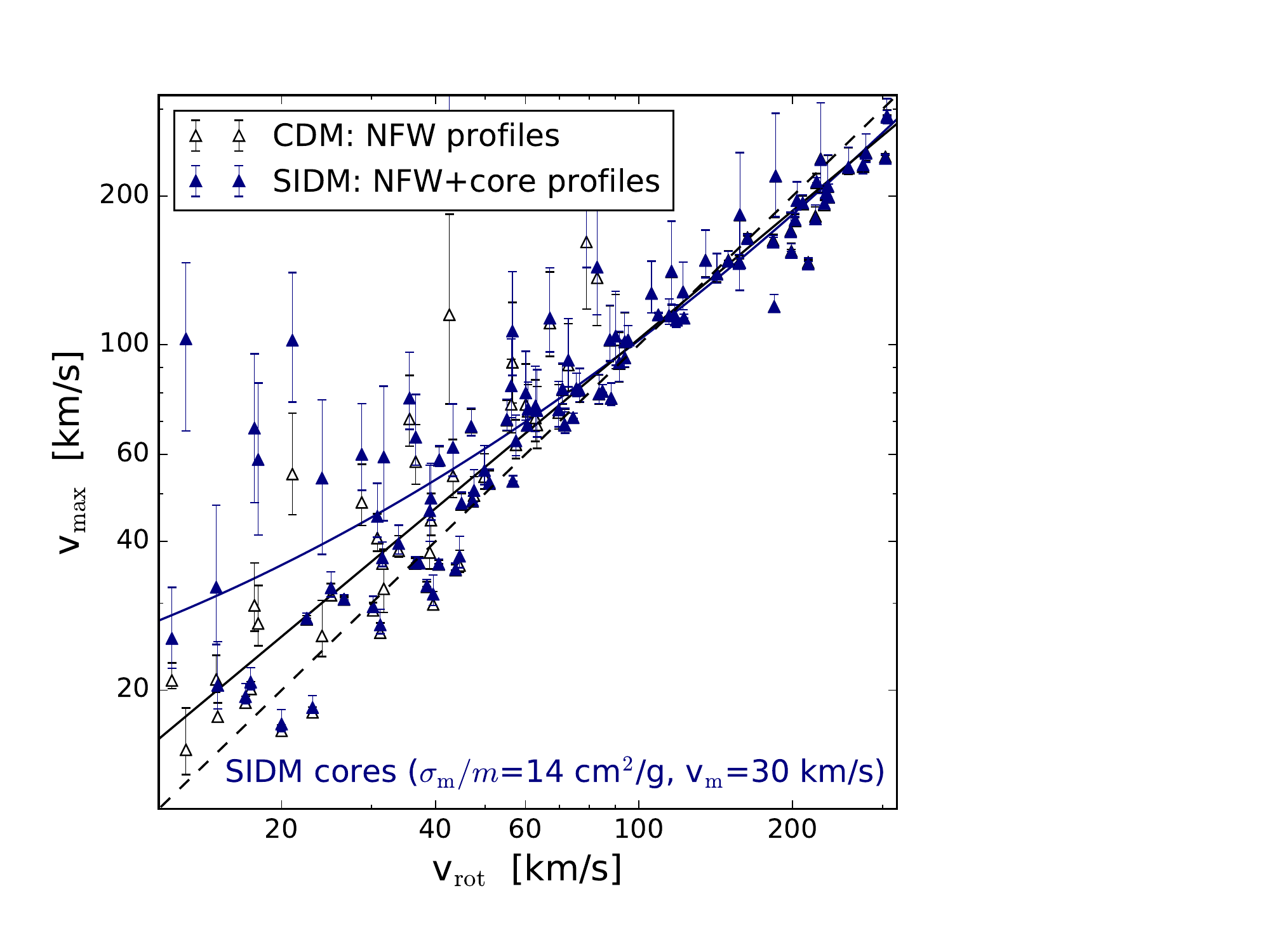}
\includegraphics[width=.328\textwidth,trim={0.6cm 0.7cm 5.57cm 0.4cm}]{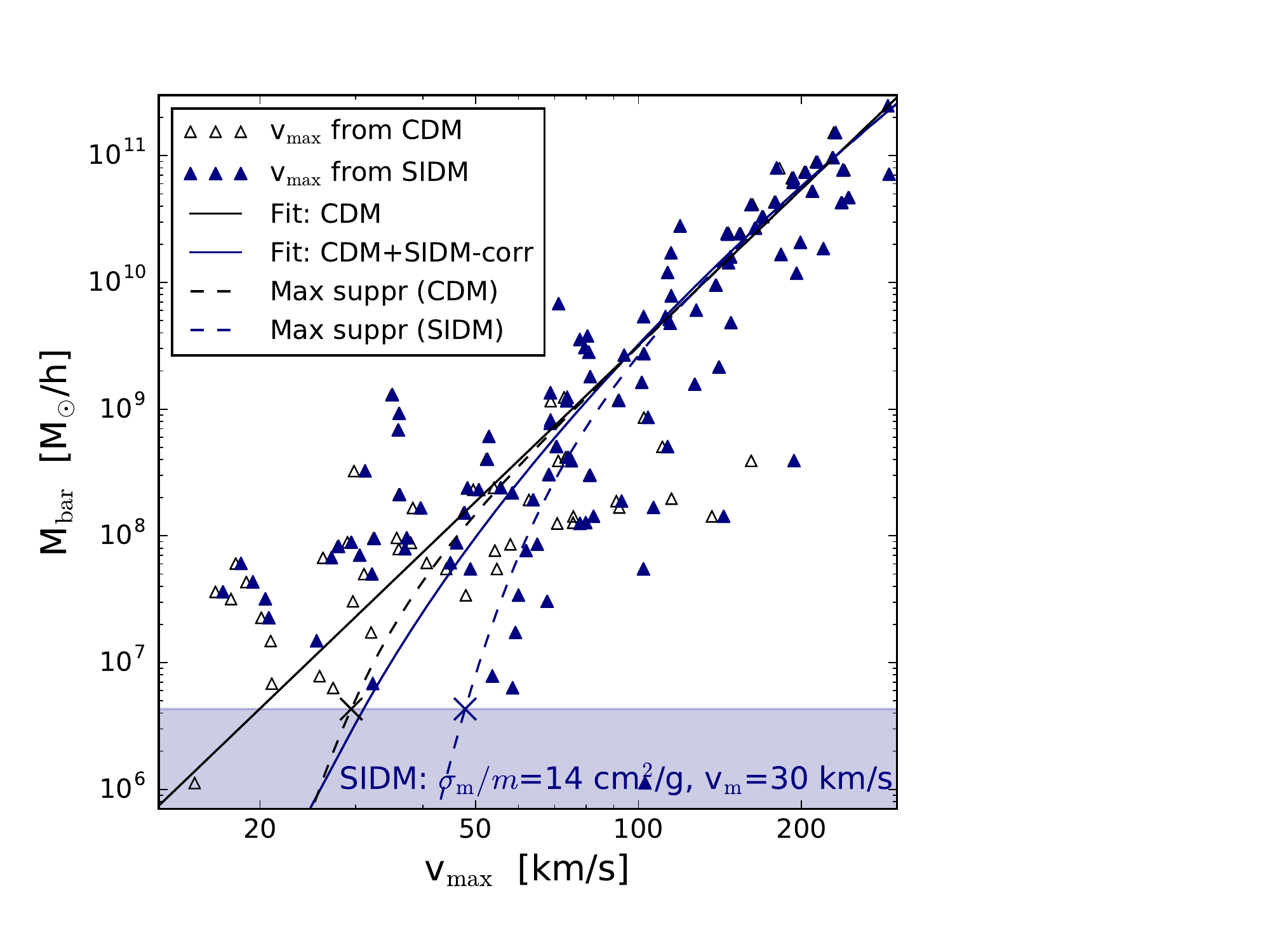}
\caption{\label{fig:SIDMprofiles}\emph{Left:} NFW velocity profiles (assuming average concentrations) for CDM (black lines) and SIDM (dark-blue lines) fitted to the observed circular velocities of the selected galaxy sample (black symbols with error-bars). \emph{Middle:} Relation between $v_{\rm max}$ and $v_{\rm rot}$ based on the fits from the left panel (empty triangle for CDM, full triangles for SIDM). The error-bars indicate the change of $v_{\rm max}$ when the halo concentrations are varied by 1-$\sigma$ around the mean. The black and brown lines are linear and quadratic least square fits, respectively. The rotational velocity ($v_{\rm rot}$) is defined via HI line width ($w_{50}$) and galaxy inclination ($i$), i.e. $v_{\rm rot}\equiv w_{50}/(2\sin i)$. \emph{Right:} Baryonic Tully-Fischer (BTF) relation based on $v_{\rm max}$ from CDM and SIDM (empty and full triangles). The solid black line is a linear fit to the BTF relation for CDM. The dark-blue line is obtained from the black line by replacing the mean values of $v_{\rm max}$ from CDM with the ones from SIDM (see lines in middle panel). The dashed lines show the strongest allowed suppression of the BTF (3-$\sigma$ excluded by the data), while the crosses indicate the velocities where these lines cross the K13 survey limit (grey band). The zero baryon suppression (solid lines) and the maximum baryon suppression (dashed lines) represent the allowable range of galaxy formation models that we will later use in calculating the VF.}}
\end{figure*}

\section{Self-Interacting dark matter}\label{sec:SIDM}
The concept of self-interacting cold dark matter (SIDM) became popular after \citet{Spergel:1999mh} showed that it could provide a better match to dwarf galaxy observations than the standard CDM model. However, it was soon realised that strong self-interactions are in conflict with observations at the scale of galaxy clusters, thereby ruling out the most simple SIDM scenarios \citep{Yoshida:2000uw,MiraldaEscude:2000qt}. More recently, the SIDM model regained popularity thanks to the realisation that previous limits were set too stringently \citep{Rocha:2012jg,Peter:2012jh}, and that velocity-dependent SIDM models easily evade limits from clusters while being well motivated by particle physics \citep{Feng:2009mn,Feng:2010zp,Loeb:2010gj}. In addition, some observational studies based on strong-lensing found offsets between the mass centres of the stellar and the DM components in clusters, which could be explained by SIDM models \citep{Williams:2011pm,Massey:2015dkw}.

Concerning the velocity function (VF), only SIDM with velocity-dependent cross sections has the potential to reduce the discrepancy on small scales without modifying the large scales. All velocity-independent models alter both small and large scales and can therefore be discarded as a solution to the observed discrepancy of the VF. For the velocity-dependent cross section, we follow \citep{Feng:2010zp,Loeb:2010gj} and assume a Yukawa force interaction leading to a scattering cross section
\begin{equation}\label{vdCS}
  \frac{\sigma}{\sigma_{\rm m}}\simeq \left\{ 
    \begin{array}{ll}
      \frac{4\pi}{22.7} \beta^2\ln(1+\beta^{-1}) & \beta< 0.1 \\
      \frac{8\pi}{22.7} \beta^2(1+1.5\beta^{1.65})^{-1} & 0.1< \beta< 10^3 \\
      \frac{\pi}{22.7} (\ln\beta+1- \frac{1}{2}\ln^{-1}\beta)^2 & \beta> 10^3
    \end{array}
\right.
\end{equation}
where $m$ is the mass of the force carrier and $\beta\equiv\pi v_{\rm m}^2/v^2$. The SIDM model has two free parameters given by $(\sigma_{\rm m}/m)$ and $v_{\rm m}$.

In Fig.~\ref{fig:crosssections} we show how the SIDM cross sections depend on velocity for the models studied in this paper. All cross sections are largest at low particle velocities and strongly reduced at large velocities, showing their potential to simultaneously produce significant cores for dwarf galaxies while evading galaxy cluster constraints. The light grey band in Fig.~\ref{fig:crosssections} shows the region of parameter space where the TBTF problem is potentially alleviated due to the reduced central densities of haloes. The dark grey band indicates the region where the largest halo cores are expected. Above this scale, core-collapse starts to dominate, effectively reducing the core sizes despite even larger cross sections \citep[see][]{Elbert:2014bma}.

The cross-sections from Eq.~\ref{vdCS} have become the standard prescription for velocity-dependent SIDM in the literature. However, there are other models with shallower velocity dependence that are equally justified from a particle physics point-of-view \citep[see][]{Kaplinghat:2015aga}.

\subsection{From cross sections to halo profiles}
The most striking feature of SIDM models in contrast to the CDM scenario is the flattening of the inner part of DM halo density profiles, which is a result of multiple scattering processes in high density regions. In previous work, SIDM haloes were described by Burkert-profiles \citep{Burkert:2000di,Zavala:2012us}. This profile provides a good fit to the inner parts of a halo but slightly deviates towards large radii \citep{Rocha:2012jg}. In this paper we use the R16 profile \citep{Read:2015sta} instead, which has the advantage of becoming an NFW profile well beyond the core radius. So far, the R16 profile has only been applied to core transformations induced by baryons, but we show in Appendix \ref{app:profile} that it also provides a very accurate fit to simulated SIDM haloes from the literature.

To assign concentrations to SIDM haloes, we use the same relation as for CDM, implicitly assuming that the rare collisions, well beyond the core radius, have a negligible effect on the profile. This assumption seems reasonable but should be tested in the future.

There is a direct relation between the cross section of SIDM and the average core size of haloes. Unfortunately, no simulation-based study has ever investigated this connection systematically. It is, however, possible to determine the approximate core size ($r_c$) analytically, using estimates of the average DM interaction rate ($\Gamma$). Assuming that a fixed number of interactions per Hubble time is required to produce a core, we obtain the relation
\begin{equation}\label{coresizeestimate}
\rho_{\rm nfw}(r_{c}|M)\frac{\langle\sigma v\rangle (r_{c})}{m}\simeq\Gamma.
\end{equation}
Here, we closely follow the approach of \citet{Dooley:2016ajo}, but we assume an NFW profile (instead of a Hernquist profile) and we use an interaction rate of $\Gamma=0.4$ Gyr$^{-1}$ (instead of $\Gamma=1.0$ Gyr$^{-1}$). The latter gives a better match to simulated SIDM haloes in combination with the R16 core definition (see the appendix for more details). Following \citet{Vogelsberger:2012ku}, the average velocity-weighted cross sections are given by the integral over the Maxwell-Boltzmann distribution, i.e.
\begin{equation}\label{velweightedcs}
\langle\sigma v\rangle(r)=\frac{1}{\sqrt{4\pi}\sigma_{\rm vel}^3(r)}\int dv v^2 (\sigma v)\exp \left[-\frac{v^2}{4\sigma_{\rm vel}^2(r)}\right].
\end{equation}
The velocity dispersion $\sigma_{\rm vel}$ depends on the halo profile and can be calculated by solving the isotropic Jeans equation, $d(\rho\sigma_{\rm vel}^2)/dr=-\rho d\phi/dr$ (where $\phi$ is the potential and $\rho$ is the DM density), leading to
\begin{equation}\label{veldisp}
\sigma_{\rm vel}^2 (r)=\frac{G}{\rho_{\rm nfw}(r)}\int_{r}^{\infty}dx\frac{M_{\rm nfw}(x)\rho_{\rm nfw}(x)}{x^2}.
\end{equation}
Eq.~(\ref{coresizeestimate}) can be combined with Eqs.~(\ref{velweightedcs}, \ref{veldisp}) to find a relation between core size ($r_c$) and halo mass ($M$). Since this relation is calibrated to SIDM simulations from the literature (via the interaction rate parameter $\Gamma$), it is expected to provide reasonably accurate results over a large range of scales. A similar approach to estimate halo cores from SIDM cross sections can be found in \citet{Kaplinghat:2015aga}. 

\begin{figure*}
\center{
\includegraphics[width=.328\textwidth,trim={0.6cm 0.7cm 4.67cm 0.4cm}]{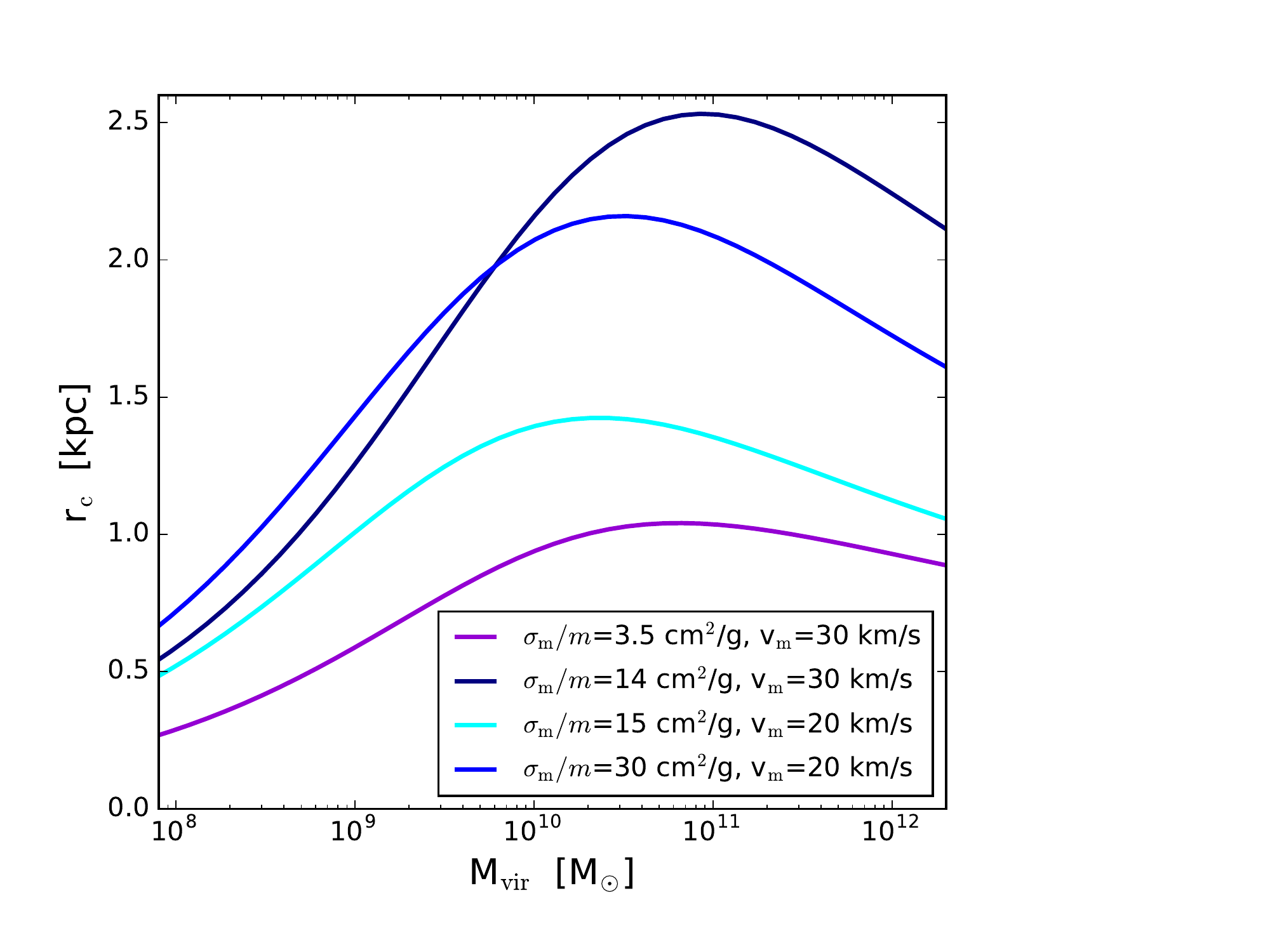}
\includegraphics[width=.328\textwidth,trim={0.6cm 0.7cm 4.67cm 0.4cm}]{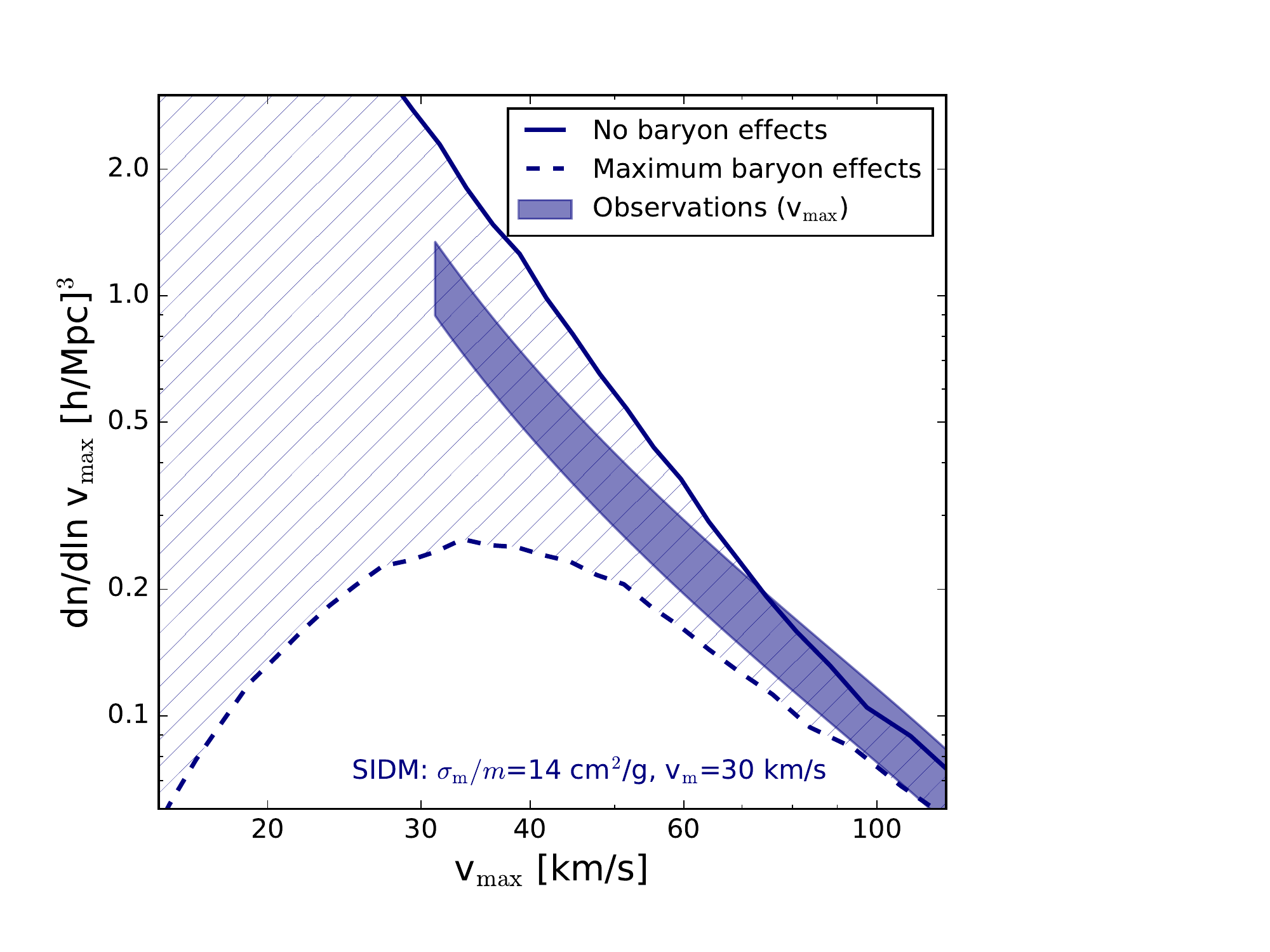}
\includegraphics[width=.328\textwidth,trim={0.6cm 0.7cm 4.67cm 0.4cm}]{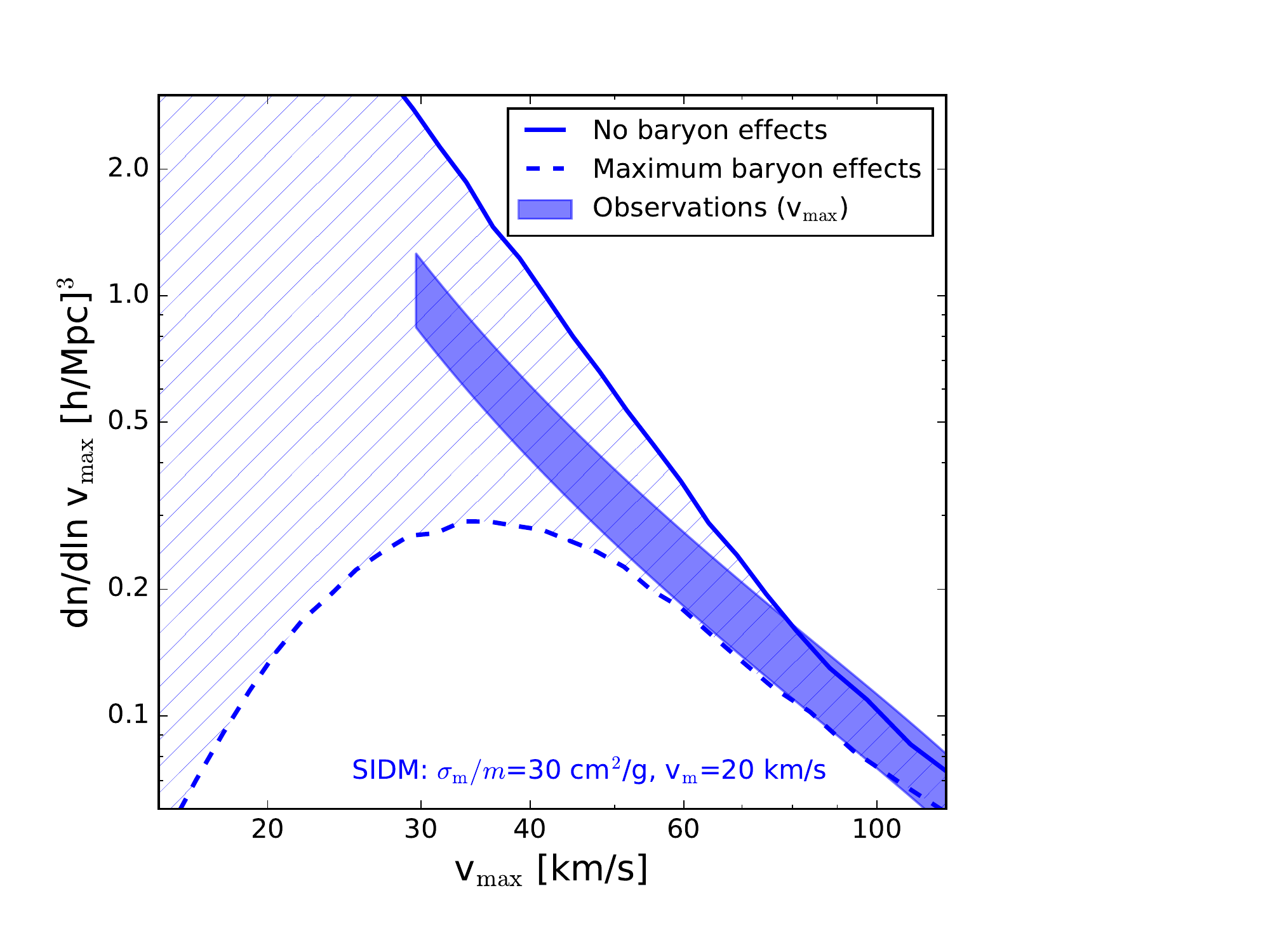}\\
\includegraphics[width=.328\textwidth,trim={0.6cm 0.7cm 4.67cm 0.4cm}]{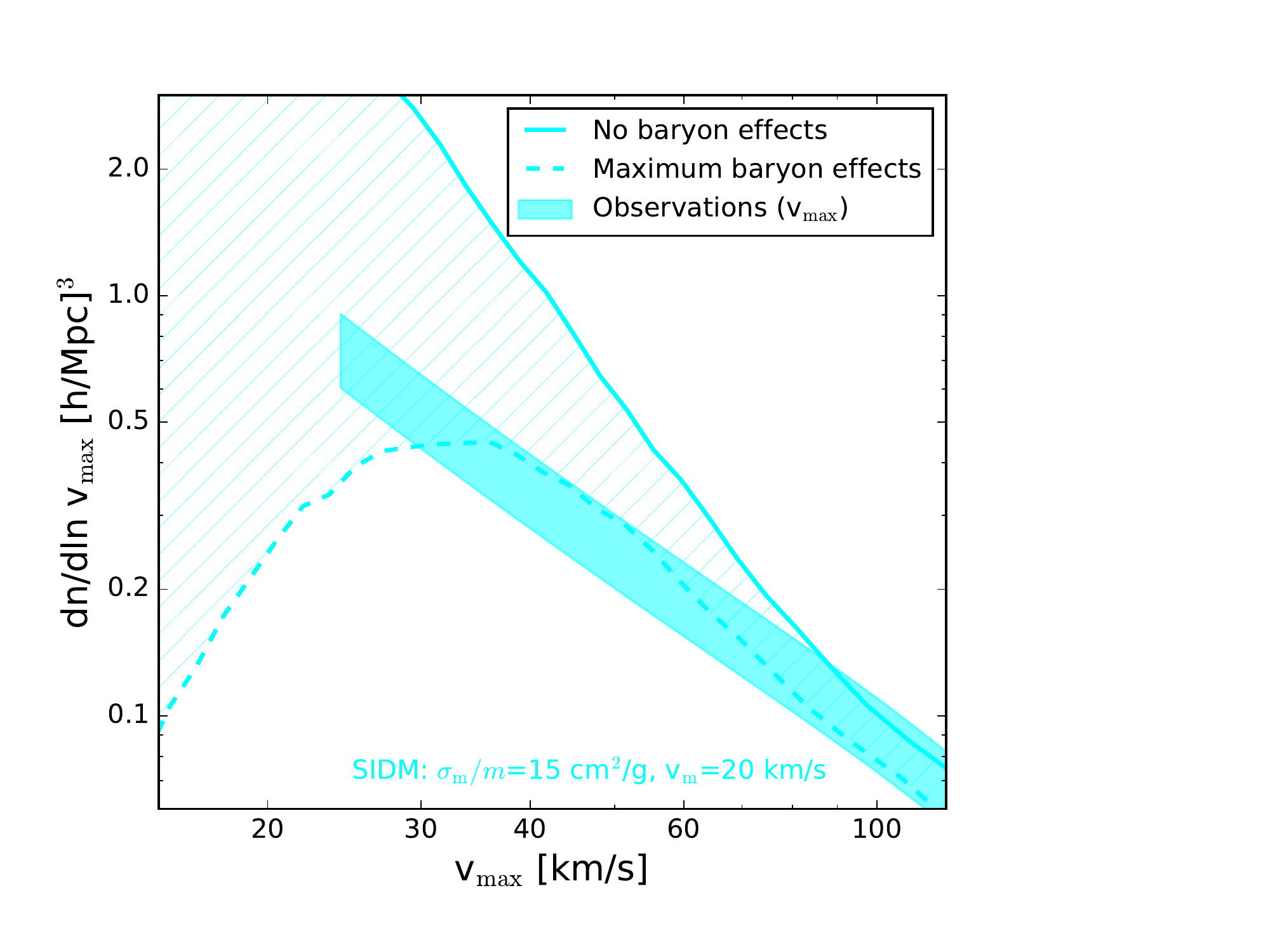}
\includegraphics[width=.328\textwidth,trim={0.6cm 0.7cm 4.67cm 0.4cm}]{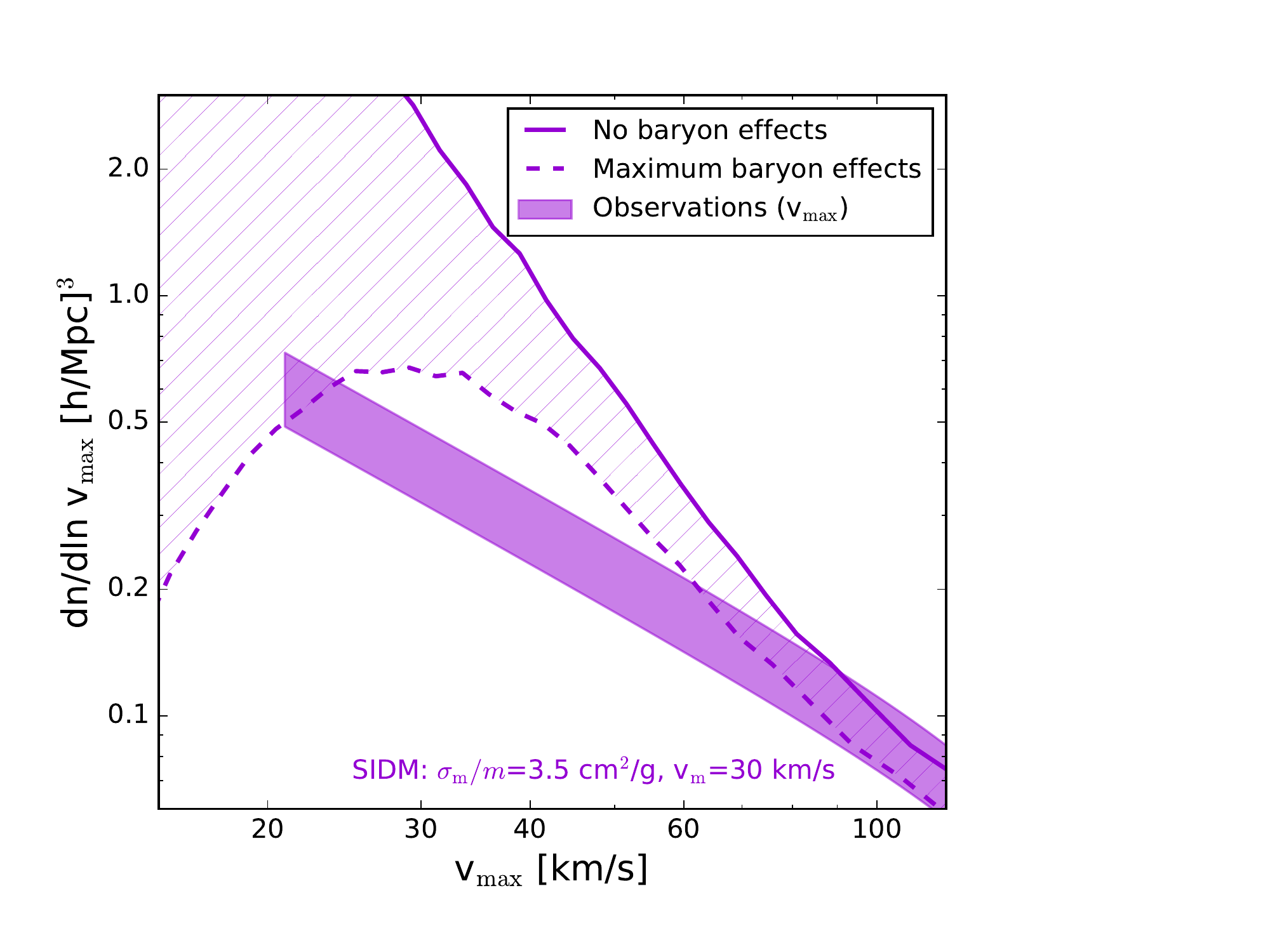}
\includegraphics[width=.328\textwidth,trim={0.6cm 0.7cm 4.67cm 0.4cm}]{Figs/velfct_CDM.pdf}
\caption{\label{fig:SIDMvelfct}The effective core sizes (top left) and the resulting velocity functions with respect to $v_{\rm max}$ for self-interacting DM scenarios. The CDM model is shown on the bottom right for comparison. The hatched area between solid and dashed lines illustrate the theoretical uncertainty related to unknown baryonic processes (see Sec.~\ref{sec:maxbaryoneffects}). The shaded bands represent the observed velocity functions corrected for $v_{\rm max}$. Cores above $r_c\sim1-1.5$ kpc are required to solve the mismatch between observations and theory.}}
\end{figure*}

In the top-left panel of Fig.~\ref{fig:SIDMvelfct}, we plot the relation between core size ($r_c$) and halo mass ($M$) for the SIDM models studied in this paper. The halo cores only vary by a factor of a few over a large range of mass scales. This is in strong contrast to SIDM models with velocity-independent cross sections, where there is a strong power-law dependence with core size increasing with halo mass \citep[see][]{Dooley:2016ajo}.

Given a halo density profile (i.e. Eq.~\ref{R16profile}) and a core radius ($r_{c}$) for SIDM, we can perform profile fits to all the galaxies in the selected sample. In the left panel of Fig.~\ref{fig:SIDMprofiles} we plot the velocity profiles of CDM (grey lines) and SIDM (with $\sigma_m/m=14$ cm$^2$/g, v$_{\rm m}=30$ km/s, dark-blue lines) fitted to the observed circular velocities $v_{\rm out}$ at $r_{\rm out}$ (symbols). The large SIDM cores lead to steep velocity profiles at small radii, requiring that small galaxies inhabit larger haloes. This becomes even more evident in the middle panel of Fig.~\ref{fig:SIDMprofiles} where the maximum circular velocity ($v_{\rm max}$) is plotted against the observed rotation velocity derived from the HI line width ($v_{\rm rot}$). There is a strong flattening and an increase of scatter visible in $v_{\rm max}$-$v_{\rm rot}$ relation towards small velocity scales (fitted by the solid dark-blue line). The latter is in line with the recently predicted higher variability of SIDM rotation curves with respect to CDM \citep{Elbert:2016dbb,Kamada:2016euw,Creasey:2016aaa}\footnote{The larger scatter of SIDM rotation curves is the result of an interplay between the stellar and the DM components. Depending on the number and distribution of stars in the halo centres, the DM component may or may not experience core collapse, resulting in a large diversity of profiles. In principle, this effect is testable by combining HI rotation curves with the observed stellar density profiles.}

Finally, we plot the $v_{\rm max}$-$M_{\rm bar}$ relation for SIDM in the right panel of Fig.~\ref{fig:SIDMprofiles}. The relation is identical to the one of CDM at large velocity scales and shows both a downturn and an increase of scatter towards smaller scales. Following the approach used for the warm and mixed DM models, we define the function $\mathcal{M}(v_{\rm max})$ by modifying the linear fit from CDM (using the average shift between $v_{\rm max}$ from SIDM and CDM, i.e. the vertical separation between the black and dark-blue lines in the middle panel). This leads to the solid blue line in the right panel of Fig.~\ref{fig:SIDMprofiles}. The model with maximum baryon suppression, $\mathcal{M}_{\rm supp}(v_{\rm max})$,  is shown as a dashed blue line. The line crosses the completeness limit of the K13 sample at the characteristic velocity, $v_c$ (dark-blue cross). The value of $v_c$ is model dependent (growing for increasing cross sections) and sets the maximum scale at which the theoretical abundance of galaxies hosted by SIDM haloes could be affected by suppression effects from baryonic processes.

Note that only one particular SIDM model is illustrated in Fig.~\ref{fig:SIDMprofiles} for brevity. Other scenarios show similar trends with growing differences between SIDM and CDM for larger particle cross-sections.

\subsection{Velocity function}
In contrast to warm and mixed dark matter, the self-interacting DM model does not lead to a reduction of the halo abundance but simply modifies their inner structure. Obviously, the disagreement between the observed and predicted velocity function of galaxies can only be solved with substantially larger cores than the ones induced by baryons, which we showed in Sec.~\ref{sec:vmax} to be not large enough to affect the result.

In Fig.~\ref{fig:SIDMvelfct} we plot the VF of the four self-interacting DM scenarios introduced above. The two models with large cross sections and cores above $r_c\sim2$ kpc are able to fully reconcile theory with observations (see top panels). The third model is also marginally consistent with observations while the last model only slightly reduces the tension with respect to CDM. 

In general, the SIDM models lead to a steepening of the observed $v_{\rm max}$-VF, while the predicted VF does not become shallower (as is the case for WDM and MDM). However, the VF can be more strongly suppressed by baryon processes in SIDM as opposed to CDM. This is due to the shallower $v_{\rm max}$-$v_{\rm rot}$ relation and the increased scatter, allowing for a stronger downturn of the $v_{\rm max}$-BTF relation.

While it is possible to solve the ``missing dwarfs" problem with SIDM models, sufficiently strong cross-sections are required, producing cores of $r_c\sim 1-1.5$ kpc or larger (at the relevant dwarf-galaxy scales). It remains to be established whether these models are in agreement with other potential constraints from structure formation. For example, larger cross sections may lead to faster evaporation of substructures (due to high-velocity encounters with particles from the host halo) which could destroy too many satellites in MW sized objects. Detailed simulations are required to refute or confirm these concerns.

\section{Other dark matter candidates}\label{sec:othercandidates}
In addition to the dark matter models presented in this paper, there are other scenarios with the potential to alleviate the mismatch between the predicted and observed velocity function. We will now mention some of them, however without providing a quantitative analysis.

An obvious DM candidate with the potential to solve the discrepancy are axion-like particles (ALPs). At very low mass scales of $m\sim10^{-23}$ eV, ALPs start to form coherent waves of astrophysical length scales, leading to a suppression of the power spectrum \citep{Hu:2000ke,Marsh:2013ywa}, and to the formation of soliton cores \citep{Schive:2014hza}. For this reason, ALPs are sometimes referred to as \emph{fuzzy} or \emph{wave} DM. Whether the soliton cores are sufficiently large to be of relevance for the VF is, however, still unclear \citep{Schive:2015kza,Hui:2016ltb}.

Another interesting scenario consists of DM particles coupled to some relativistic fluid like photons, neutrinos, or dark radiation \citep[see e.g.][]{CyrRacine:2012fz,Boehm:2014vja,Bringmann:2016ilk}. Such a coupling generates a suppression of the power spectrum usually combined with acoustic oscillations at the suppression scale. In terms of effects on the galaxy abundance, a similar behaviour than for WDM or MDM can be expected. It would furthermore be interesting to establish if the dark acoustic oscillation could still be visible in the galaxy velocity function \citep[see also][]{Buckley:2014hja}.

An interesting new framework to systematically study interactions of the dark sector has recently been presented under the name of \emph{ETHOS} \citep[effective theory of structure formation,][]{Cyr-Racine:2015ihg,Vogelsberger:2015gpr}. Some of the example cases investigated by the authors show models with both suppressed power spectra and significant halo cores. The VF of the local volume could provide an ideal testbed to further study such scenarios.


\section{Conclusions}\label{sec:conclusions}
In the last few years it became evident that the observed \emph{HI velocity width function} of galaxies in the local volume is in tension with predictions from gravity-only simulations based on the standard model of $\Lambda$CDM \citep[e.g. ][]{Zavala:2009ms,Papastergis:2011xe,Klypin:2014ira}. The discrepancy cannot be fully solved with baryon effects, such as strong UV photoevaporation or supernova feedback, because these processes induce a downturn in the $v_{\rm max}$-$M_{\rm bar}$ relation, which is not observed \citep[][TG16]{Trujillo-Gomez:2016pix}. This crucial point is further highlighted in Appendix \ref{app:comparison}, where we compare to other work based on abundance matching techniques, simulations, and direct mass estimates from observed dwarf galaxies.

In principle, a number of observational effects could be affecting the kinematic analysis performed in this article, and could therefore alter our conclusions regarding the viability of $\Lambda$CDM\footnote{For a detailed discussion about potential systematics regarding the galaxy sample, see Sec.~4.1 of \citet{Papastergis:2016aaa}.}. For example, the subsample of selected galaxies (with spatially resolved kinematic measurements) used to determine $v_{\max}$ could be biased with respect to the full sample used for the VF. This possibility was investigated by TG16, who found no systematic differences between the selected and full samples of local volume galaxies.  Another possibility is a bias from inaccurate estimates of galaxy inclinations. Indeed, highly inclined galaxies (which give more accurate measurements of $v_{\rm rot}$) tend to have somewhat smaller velocities at a given baryonic mass. This means, however, that reducing inclination errors would shift the observed velocity function of $v_{\rm max}$ further away from the CDM predictions, worsening the discrepancy. 

A further potential source of error are dwarf galaxies with an extremely low surface-brightness falling below the survey detection limit. If such objects exist in large numbers, they could potentially explain the difference between the observed and predicted velocity function. A detailed discussion of this possibility is given by \citet{Klypin:2014ira}. There it is argued that while some of the smallest galaxies with $v_{\rm rot}<20$ km/s could potentially stay undetected because of extremely low surface-brightness, they are very unlikely to make up a sizeable fraction of the full population\footnote{Furthermore, the most recent searches for extremely low surface brightness dwarfs around massive spirals (including Local Group dwarfs and MW and M31 satellites) find very few objects with $\mu>27$ mag/arcsec$^2$ brighter than $M_V=-10$ \citep[see e.g.][]{Merritt:2014rza}. Therefore, most of the dwarfs in the local volume that went undetected due to their low surface brightness should be below the magnitude limit of the catalogue from \citet{Karachentsev:2013ipr}, and hence would not affect our conclusions.}.

Finally, the mismatch between the observed and predicted $v_{\rm max}$-VF could originate from errors in the fitting procedure used to determine $v_{\rm max}$, with the main concern being the possibility of inner DM cores induced by stellar feedback. However, existing estimates produced by baryonic feedback effects \citep{Read:2015sta,DiCintio:2013qxa} have been shown to be too small to produce a significant effect in galaxies with kinematic measurements at large radii (see Fig.~\ref{fig:intro} and TG16). Finally, the results of this paper depend on the customary assumption that spatially resolved HI rotation measurements (including corrections for turbulence) can be used to probe the halo potential.

A more speculative but intriguing option is that the mismatch between observed and predicted abundance of isolated galaxies points towards an alternative dark matter sector. In the present paper, we investigated different DM scenarios such as warm, mixed, and self-interacting DM, and we showed that they can be in much better agreement with observations. A more detailed summary of the results follows:
\begin{itemize}
\item The warm DM (WDM) scenario is characterised by a steep cutoff in the initial power spectrum, resulting in two important effects: reduced halo abundance, and lower concentrations. The two effects work together to flatten the predicted $v_{\rm max}$-VF. Furthermore and due to the lower concentrations, observed galaxies are expected to reside in more massive haloes, yielding a steeper relation for the observed $v_{\rm max}$-VF. As a result, observations and theory agree for WDM models with (thermal-relic) masses between $m_{\rm TH}\sim 1.8-3.5$ keV (see Fig.~\ref{fig:WDMvelfct}). This includes lukewarm DM scenarios which are cold enough to avoid the most stringent constraints from the Lyman-$\alpha$ forest. Significantly warmer models with $m_{\rm TH}\lesssim1.5$ keV are disfavoured by the observed $v_{\rm max}$-VF.
\item We also considered a simple mixed DM (MDM) scenario with varying fraction of warm to cold dark matter. This model leads to a wide range of shapes for the power spectrum making MDM an ideal testbed for structure formation. The resulting effect on the VF is qualitatively similar to the case of WDM, except that the flattening can affect a wider range of scales. For large parts of the MDM parameter space, the agreement between observations and theory is highly improved with respect to the case of CDM (see Fig.~\ref{fig:MDMvelfct}). Again, this includes models that are in agreement with Lyman-$\alpha$ limits.
\item The self-interacting DM (SIDM) scenario is qualitatively different from warm and mixed DM models in the sense that it does not yield suppressed perturbations at small scales. Instead, the inner parts of halo profiles are flattened due to repeated collisions of DM particles in high-density regions. The flattening of profiles leads to the inevitable conclusion that small galaxies should inhabit more massive haloes compared to CDM. As a result, the observed $v_{\rm max}$-VF becomes steeper for increasing cross sections. The theoretical abundance of galaxies in SIDM models is unchanged with respect to CDM as long as baryonic processes are neglected. The model including the maximum allowed baryonic suppression, on the other hand, is less constrained than for CDM due to the higher values (and the increased scatter) of $v_{\rm max}$ estimated for observed dwarf galaxies. As a result, SIDM models can fully solve the tension between the predicted and observed $v_{\rm max}$-VF as long as they form sufficiently large DM cores of $r_c\gtrsim1.5$ kpc in dwarf galaxies (see Fig.~\ref{fig:SIDMvelfct}). This is only possible for models with significant cross sections which have to be velocity-dependent to avoid constraints from galaxy clusters. Whether these models are in agreement with other small-scale observables has yet to be established.
\end{itemize}

In general, all models that either suppress perturbations at dwarf galaxy scales or flatten the inner DM halo density profiles (or a combination of both) can potentially alleviate the mismatch between predicted and observed abundance of galaxies as a function of $v_{\rm max}$. This includes many more scenarios than the ones studied here. Examples are interacting, decaying, late decoupling, or boson dark matter. A more detailed investigation of such models is postponed to future work.

Upcoming large area HI surveys, such as the {\tt APERTIF} survey with the {\tt WSRT} interferometer and the {\tt WALLABY} survey with the {\tt ASKAP} interferometer will provide large samples of dwarf galaxies with spatially resolved velocity information and improved HI sensitivity. This will make it possible to track down the remaining potential systematics related to the profile fitting procedure. The new data will furthermore allow to extend the observed velocity function to smaller scales, well below 10 km/s in velocity. This should lead to the discovery of a downturn of both the VF and the $v_{\rm max}$-$M_{\rm bar}$ relation due to the effects of photo-evaporation during reionisation. Once this effect is known to better accuracy, it will be possible to come up with highly improved constraints for the particle nature of dark matter.



\section*{Acknowledgements}
We thank Anatoly Klypin for very helpful suggestions on how to improve the present manuscript.
AS acknowledges support from the Swiss National Science Foundation (PZ00P2\_161363).
EP is supported by a postdoctoral fellowship of the Netherlands Research School for Astronomy (NOVA).

\appendix

\section{Comparison with other studies}\label{app:comparison}
Several recent papers have reported solutions to the overabundance problem of the galactic velocity function within the standard model of $\Lambda$CDM. In this section we discuss how these papers differ from our work and why we think they do not remove the problem.

\begin{figure*}
\center{
\includegraphics[width=.48\textwidth,trim={0.6cm 1.0cm 5.4cm 0.8cm}]{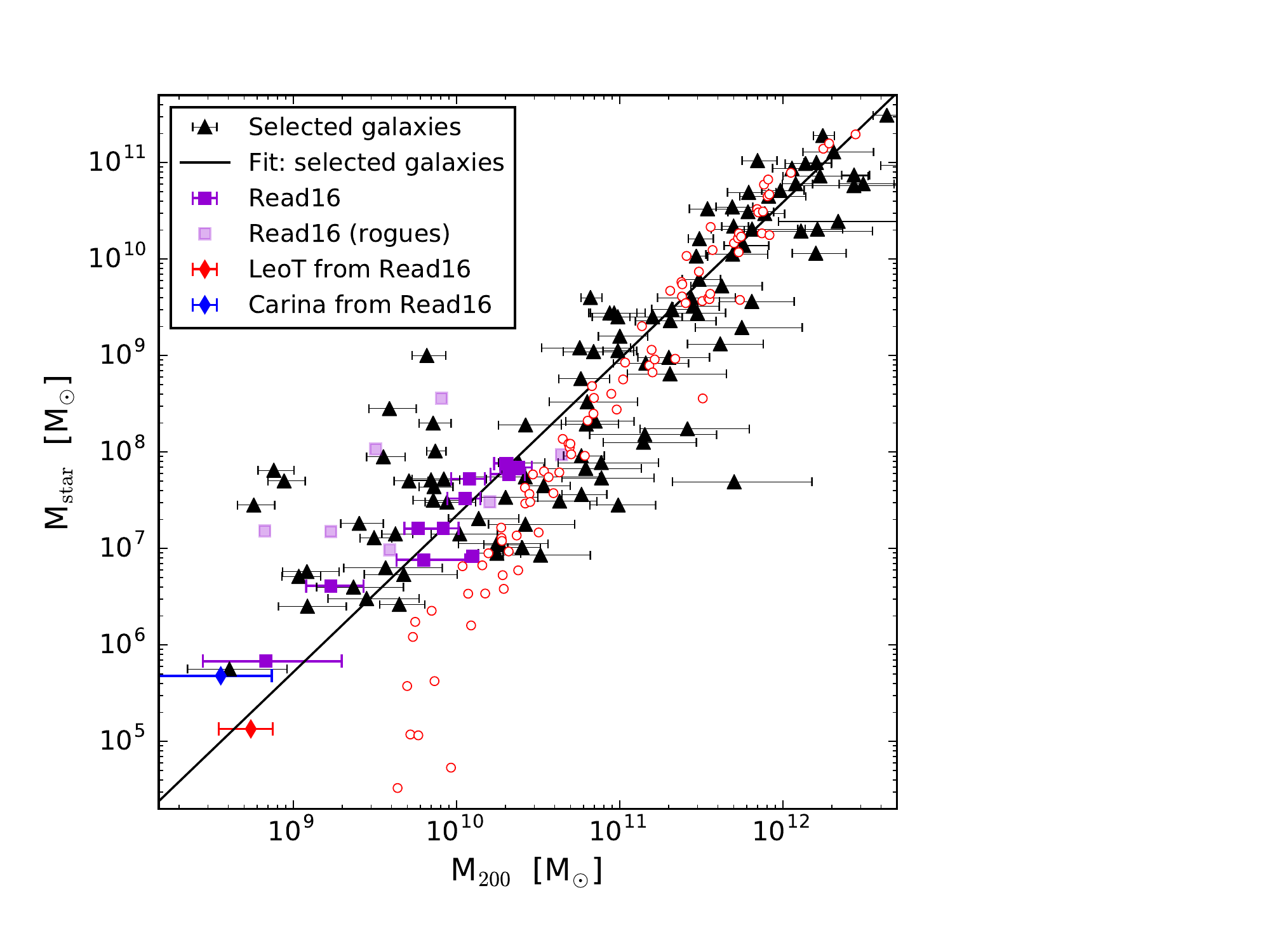}
\includegraphics[width=.48\textwidth,trim={0.6cm 1.0cm 5.4cm 0.8cm}]{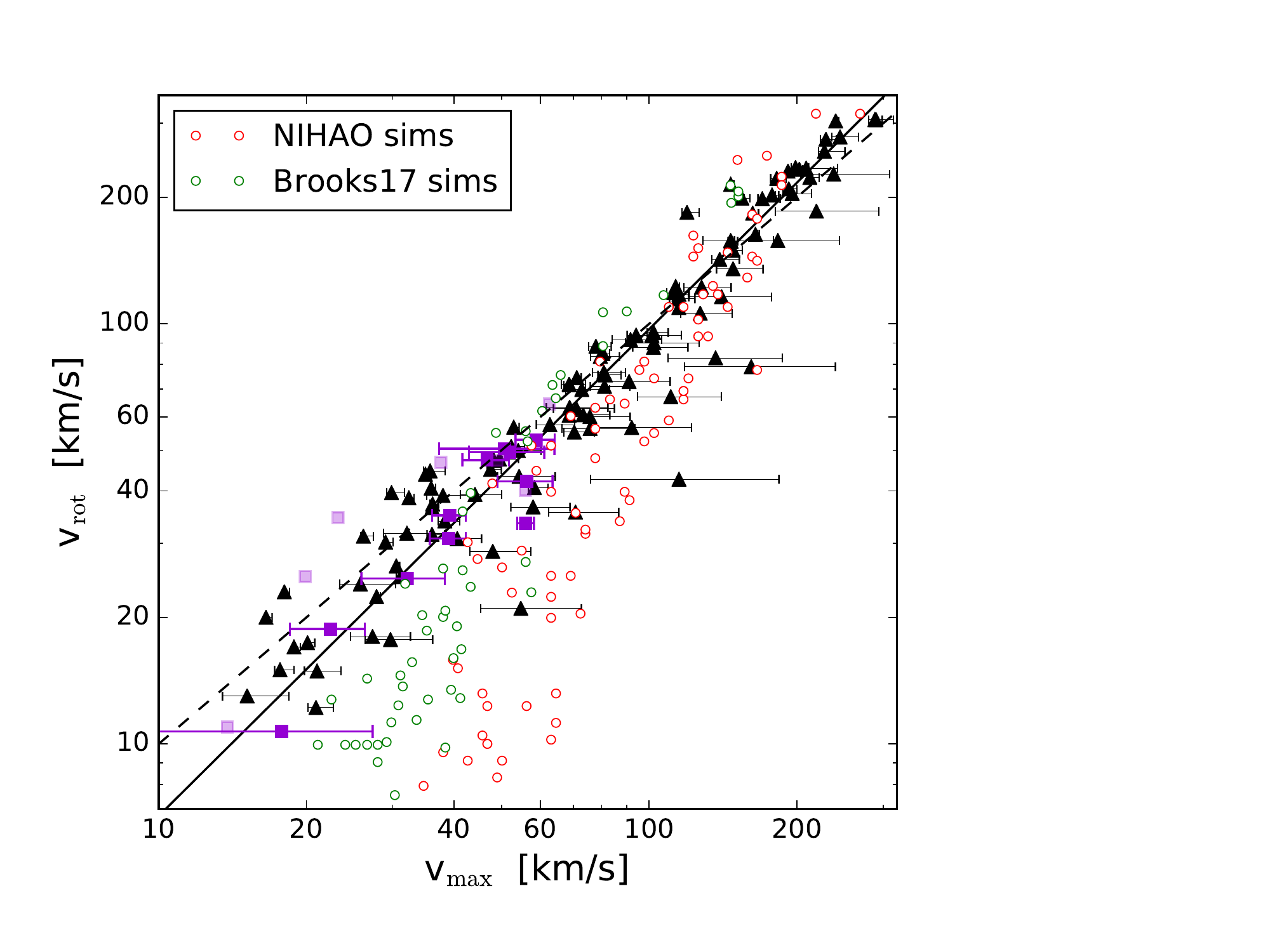}
\caption{\label{fig:appendix1}Relation between $M_{\rm star}$ and $M_{200}$ (left panel) and between $v_{\rm rot}$ and $v_{\rm max}$ (right panel) for the selected galaxy sample (black triangles) compared to individual field dwarfs from Read16 (purple symbols). The bold purple symbols correspond to reliable rotation curves while the shaded purple symbols (without error-bars) have potential systematics due to low inclinations or signs of disequilibrium (dubbed `rogues' in Read16). The red and blue data points represent mass estimates for the LeoT dwarf and the Carina satellite (Read16). Next to the direct estimates from observations, we show results from the NIHAO and the Brooks17 simulations (red and green circles) which agree at large scales but deviate towards smaller scales most relevant for the velocity function.}}
\end{figure*}

\citet{Brook:2015ofa} performed a detailed investigation of the galactic velocity function for three cases, CDM with baryonic cores, as well as one model of WDM and SIDM both without baryonic cores. Starting with the halo mass function, they use abundance matching (AM) to obtain stellar masses, from which they estimate the extent of stellar and HI discs as well as baryon induced DM cores. This allows them to obtain mock velocity profiles for each halo mass. Based on these velocity profiles, \citet{Brook:2015ofa} determine the radius where the circular velocity equals the observed $v_{\rm rot}$ from small galaxies (which leads to an empirical relation between $M_{\rm star}$ and $r_{\rm rot}$). They find this radius to be much smaller than the radius of maximum circular velocity, concluding that $v_{\rm rot}$ is considerably smaller than $v_{\rm max}$. This bias between $v_{\rm rot}$ and $v_{\rm max}$ strongly reduces the initial discrepancy between theory and observation in the VF without, however, completely solving it. \citet{Brook:2015ofa} show that the remaining tension can be solved by either assuming baryon-induced cores within $\Lambda$CDM or alternative dark matter (i.e. WDM or SIDM without baryonic cores).

The main weakness of the method applied by \citet{Brook:2015ofa} is that it relies on abundance matching. The AM technique matches observations to the $\Lambda$CDM model by \emph{assigning} small galaxies to very large haloes, without \emph{testing} this assignment against observations. As a consequence, \citet{Brook:2015ofa} obtain small HI radii (via their empirical $M_{\rm star}$-$r_{\rm rot}$ scaling relation) and therefore small values for $v_{\rm rot}$, which largely alleviates the initial discrepancy between the predicted and observed VF.

Very recently, \citet{Brook:2015eva} published a paper, where abundance matching combined with the observed baryonic Tully-Fisher (BTF) relation is used to estimate the bias between $v_{\rm rot}$ and $v_{\rm max}$. They find that the discrepancy between observed and predicted velocity function disappears entirely, provided the adequate Tully-Fisher relation is used. While this paper reveals systematical differences between different observations and highlights the importance of an adequate velocity definition, it does not provide a test for $\Lambda$CDM on its own\footnote{Indeed, for the ideal case where the BTF, the VF, as well as the AM relation is based on one single set of observations, the discrepancy of the VF has to disappear by construction for a large number of different cosmological models.}. A further necessary requirement for testing the underlying cosmological model is to verify the applied AM relation.

Attempts to verify abundance matching of $\Lambda$CDM by estimating the halo mass of observed galaxies have been performed by several papers in the past. \citet{Papastergis:2014aba} used kinematical information from the extended HI content of dwarf galaxies to show that there is a discrepancy between the data and AM relations from the local volume, irrespectively of whether baryon-induced cores are assumed or not. More recently, \citet{Pace:2016oim} estimated halo masses based on careful analysis of full HI rotation curves for a few field galaxies, obtaining similar results. \citet{Karukes:2016eiz} also find a discrepancy with AM expectations based on $\Lambda$CDM, by analysing the rotation curves of a sample of 36 late-type dwarfs. \citet{Brook:2014hda}, on the other hand, used stellar kinematics to estimate halo masses of dwarf galaxies. They find better agreement with AM relations from the local volume, mainly due to their assumption of cored profiles from baryon processes. However, stellar kinematics only probe the very inner region of haloes, which are subject to large uncertainties and potential systematics in the halo mass estimates. Indeed, small differences in the model are amplified leading to large differences in halo mass.

Recently, \citet{Maccio:2016egb} published a study on the VF based on the NIHAO\footnote{The Acronym NIHAO stands for Numerical Investigation of a Hundred Astrophysical Objects.} suite of hydrodynamical simulations with full metal cooling and standard recipes for sub-grid effects such as star formation and supernova feedback. They report a very large bias between $w_{50}$ and $v_{\rm max}$ fully solving the discrepancy between the observed and predicted VF.
There are two main reasons why the results from \citet{Maccio:2016egb} differ from ours. First, the HI content of the NIHAO galaxies is less extended than the one from the selected sample of observed galaxies leading to smaller values of $w_{50}$ (or $v_{\rm rot}$) compared to $v_{\rm max}$ (see TG16 and \citet{Papastergis:2016aaa} for a detailed comparison). Second, the strong feedback effects present in the NIHAO simulations make the NIHAO galaxies reside in very massive haloes compared to the mass estimates from the selected sample.

Finally, during the review process of this work, \citet[][henceforth Brooks17]{Brooks:2017rfe} published a study based on a suite of hydrodynamical simulations that claims to fully solve the apparent discrepancy of the velocity function. Similarly to \citet{Maccio:2016egb}, they obtain a more significant bias between $v_{\rm rot}$ and $v_{\rm max}$ compared to what we find in our analysis. A closer look at their results reveals that they are able to completely close the gap between the observed and predicted VF between $v_{\rm rot}=20-50$. However, there is some remaining discrepancy at both larger and smaller velocities, which they correct in the latter case by assuming a cutoff due to reionisation.

As a result of very efficient feedback recipes, both the NIHAO and the Brooks17 simulations obtain larger halo masses and larger maximum circular velocities than what we find by analysing HI kinematics of dwarf galaxies. This is illustrated in Fig.~\ref{fig:appendix1}, where the left and right panels show the $M_{\rm star}-M_{\rm 200}$ and the $v_{\rm rot}-v_{\rm max}$ relations, respectively. For both relations the selected galaxy sample (black symbols with error bars representing the dependence on the concentration parameter) is well described by a power law (black solid lines) while the NIHAO and Brooks17 simulations (red and green circles) exhibit a downturn towards small mass and velocity scales. This is a direct consequence of their strong feedback recipes which reduce the amount of stars and gas in a halo of a given size.

In Fig.~\ref{fig:appendix1} we also compare our results to independent estimates of halo mass and maximum circular velocity from \citet[][henceforth Read16]{Read:2016aaa}. They used fully resolved rotation curves of individual field dwarfs accounting for stellar and gaseous components as well as baryon-induced cores. The results of Read16 are shown as purple squares, where bright symbols represent dwarfs with reliable rotation curves while the shaded symbols without error-bars denote data with potential systematics from inclination or signs of disequilibrium (dubbed `rogues' in Read16). It is very encouraging that the results from Read16 agree well with our own mass estimates.

In summary, there is a systematic difference between the halo mass of simulated galaxies (from the NIHAO or the Brooks2017 simulations) and direct mass estimates from local field dwarfs which could point towards a genuine problem of hydrodynamical simulations at dwarf galaxy scales. However, we want to point out that the current observational data is too sparse to support any strong conclusions. Upcoming large area HI surveys with interferometric data will highly improve the observational situation in the next few years.

\begin{figure*}
\center{
\includegraphics[width=.48\textwidth,trim={0.6cm 0.9cm 3.5cm 0.9cm}]{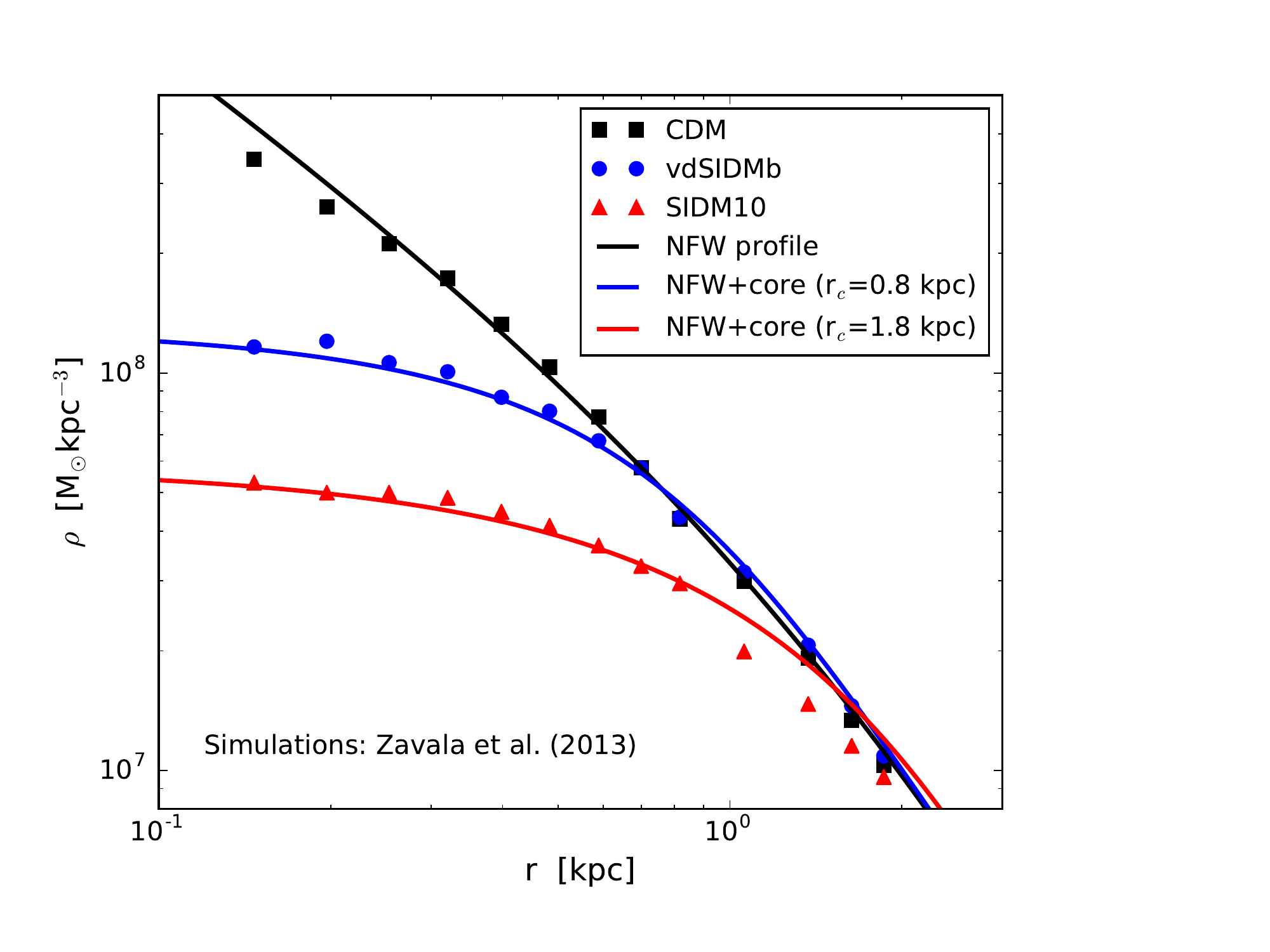}
\includegraphics[width=.48\textwidth,trim={0.6cm 0.9cm 3.5cm 0.9cm}]{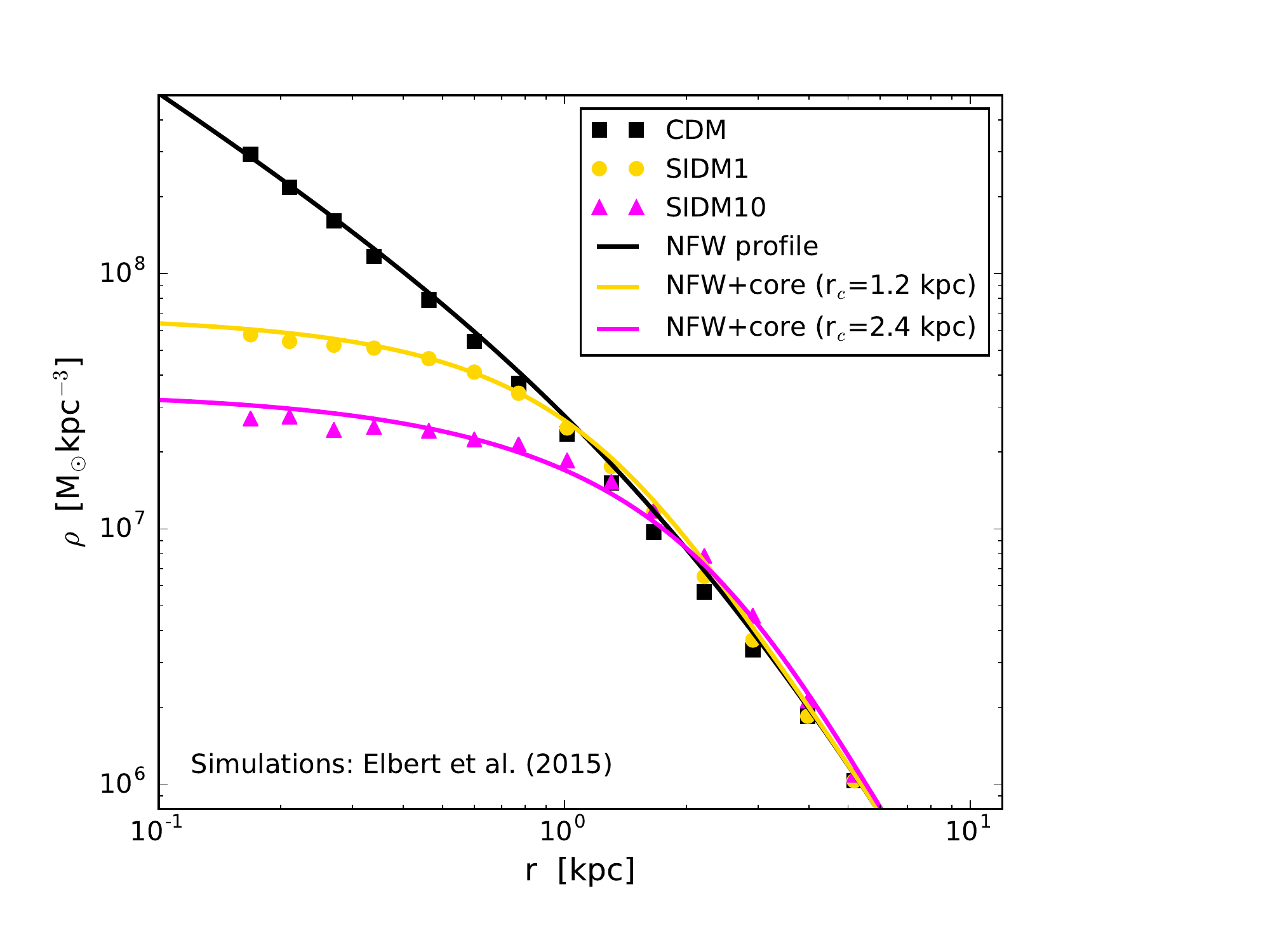}
\caption{\label{fig:appendix2}Simulated profiles of dwarf galaxy haloes from the literature \citep{Zavala:2012us,Elbert:2014bma} fitted with the R16 profile (coloured dots and lines, respectively). \emph{Left:} The model vdSIDMb (blue) corresponds to velocity dependent SIDM (with $\sigma_{\rm m}/m=35$ cm$^2$/g and $v_{\rm m}=10$ km/s) while SIDM10 (red) refers to a velocity independent model with $\sigma/m=10$ cm$^2$/g. \emph{Right:} Two models with velocity independent cross section of $\sigma/m=1$ cm$^2$/g (yellow) and $\sigma/m=10$ cm$^2$/g (magenta).}}
\end{figure*}

\section{A new halo density profile for SIDM}\label{app:profile}
Here we demonstrate that the R16 profile \citep{Read:2015sta} is not only suitable to describe cores from baryonic feedback, but also provides very accurate fits to profiles of self-interacting dark matter (SIDM) haloes. In Eq.~(\ref{R16profile}) we introduced the R16 mass profile which consists of the NFW mass profile multiplied with a simple two-parameter function. The density profile can be obtained from Eq.~(\ref{R16profile}) by a simple derivative, i.e.,
\begin{equation}
\rho_{\rm R16}(r) = \rho_{\rm nfw}(r)f^n+\frac{n f^{(n-1)}(1-f^2)}{4\pi r^2r_c}M_{\rm nfw}(r),
\end{equation}
where $n=1$ and $f(r)$ is given by Eq.~(\ref{R16profile}). The R16 profile has the advantage of converging to the NFW profile for $r\gg r_c$, where DM self-interactions are negligible.

In Fig.~\ref{fig:appendix2} we show fits using the R16 profile to a few simulated SIDM profiles found in the literature. The left panel shows data points for two haloes from \citet{Zavala:2012us} corresponding to velocity-dependent model with $\sigma_{\rm m}/m=35$ cm$^2$/g and $v_{\rm m}=10$ km/s (blue, dSIDMb) and a velocity-independent model with $\sigma/m=10$ cm$^2$/g (red, SIDM10). The haloes are well fitted with R16 profiles with $r_c=0.8$ kpc and $r_c=1.8$ kpc. The right panel shows two haloes from  \citet{Elbert:2014bma} out of velocity-independent SIDM simulations with $\sigma/m=1$ cm$^2$/g (yellow, SIDM1) and $\sigma/m=10$ cm$^2$/g (magenta, SIDM10), respectively. Again the R16 profile provides an accurate fit to the simulated halo profiles. The core sizes for these haloes are $r_c=1.2$ kpc and $r_c=2.4$ kpc.

\bibliographystyle{mnras}
\bibliography{ASbib}

\end{document}